\documentclass[referee,onecolumn]{aa}

\usepackage{natbib,twoopt}
\bibpunct{(}{)}{;}{a}{}{,} 
\makeatletter
\newcommandtwoopt{\citeads}[3][][]{\href{http://adsabs.harvard.edu/abs/#3}%
{\def\hyper@linkstart##1##2{}%
\let\hyper@linkend\@empty\citealp[#1][#2]{#3}}}
\newcommandtwoopt{\citepads}[3][][]{\href{http://adsabs.harvard.edu/abs/#3}%
{\def\hyper@linkstart##1##2{}%
\let\hyper@linkend\@empty\citep[#1][#2]{#3}}}
\newcommandtwoopt{\citetads}[3][][]{\href{http://adsabs.harvard.edu/abs/#3}%
{\def\hyper@linkstart##1##2{}%
\let\hyper@linkend\@empty\citet[#1][#2]{#3}}}
\newcommandtwoopt{\citeyearads}[3][][]%
{\href{http://adsabs.harvard.edu/abs/#3}
{\def\hyper@linkstart##1##2{}%
\let\hyper@linkend\@empty\citeyear[#1][#2]{#3}}}
\makeatother

\RequirePackage{graphicx}
\RequirePackage[varg]{txfonts}

\begin{document}

\title{Efficiency of thermal relaxation by radiative processes in protoplanetary discs: constraints on hydrodynamic turbulence} 
\author{M.~G.~Malygin\inst{1,2}, H.~Klahr\inst{1}, D.~Semenov\inst{1}, Th.~Henning\inst{1} and C.~P.~Dullemond\inst{3}}

   \institute{Max-Planck-Institut f\"ur Astronomie, K\"onigstuhl 17, D-69117 Heidelberg, Germany\\
              \email{malygin@mpia.de}
         \and
     Fellow of the International Max-Planck Research School for Astronomy and Cosmic Physics at the University of Heidelberg (IMPRS-HD)
         \and
     Zentrum f\"ur Astronomie der Universit\"at Heidelberg, Institut f\"ur Theoretische Astrophysik, Albert-Ueberle-Stra{\ss}e 2, 69120 Heidelberg, Germany
             }

   \date{Received date; accepted date}

 
  \abstract
       {
           Hydrodynamic, non-magnetic instabilities can provide turbulent stress in the regions of protoplanetary discs, where the magneto-rotational instability can not develop. 
           The induced motions influence the grain growth, from which formation of planetesimals begins. 
           Thermal relaxation of the gas constrains origins of the identified hydrodynamic sources of turbulence in discs. 
       }
       {
           We aim to estimate the radiative relaxation timescale of temperature perturbations in protoplanetary discs. 
           We study the dependence of the thermal relaxation on the perturbation wavelength, the location within the disc, the disc mass, and the dust-to-gas mass ratio. 
           We then apply thermal relaxation criteria to localise modes of the convective overstability, the vertical shear instability, and the zombie vortex instability. 
       } 
       {
           For a given temperature perturbation, we estimated two timescales: the radiative diffusion timescale $t_\mathrm{thick}$ and the optically thin emission timescale $t_\mathrm{thin}$. 
           The longest of these timescales governs the relaxation: $t_\mathrm{relax} = \max \left( t_\mathrm{thick}, t_\mathrm{thin} \right)$. 
           We additionally accounted for the collisional coupling to the emitting species. 
           Our calculations employed the latest tabulated dust and gas mean opacities.
       }
       {
           The relaxation criterion defines the bulk of a typical T Tauri disc as unstable to the development of linear hydrodynamic instabilities. 
           The midplane is unstable to the convective overstability from at most $2\mbox{ au}$ and up to $40\mbox{ au}$, as well as beyond $140\mbox{ au}$. 
           The vertical shear instability can develop between $15\mbox{ au}$ and $180\mbox{ au}$. 
       The successive generation of (zombie) vortices from a seeded noise can work within the inner $0{.}8\mbox{ au}$. 
       } 
       {
           A map of relaxation timescale constrains the origins of the identified hydrodynamic turbulence-driving mechanisms in protoplanetary discs. 
           Dynamic disc modelling with the evolution of dust and gas opacities is required to clearly localise the hydrodynamic turbulence, and especially its non-linear phase. 
       }


   \titlerunning{ Radiative relaxation in PPDs }
   \authorrunning{ M.~G.~Malygin et al. }
   \maketitle
%


\section{Introduction}\label{sec1:intro}
Despite the ample observational manifestation, accretion discs around young stars remain a puzzle. 
In such environments, the major mode of planet formation takes place \citep{JOHANSEN14,RAYMOND14,CHABRIER14}. 
The exact mechanism of turbulence generation in circumstellar discs is still debated \citep{TURNER14} and there may be multiple mechanisms operating simultaneously in different locations of a disc yet interacting with each other \citep{ARMITAGE15}. 
From the theoretical perspective, it is clear that turbulence is central both for angular momentum transport and planetesimal aggregation \citep{TURNER14}. 

One of the most promising models of turbulence in protoplanetary discs is the magneto-rotational instability \citep[MRI,][]{BALBUS91}. 
The MRI operates in weakly magnetised yet well-coupled shearing flows. 
The major localising constraint on the MRI is a low ionisation degree of the gas \citep{GAMMIE96,DZYURKEVICH13}. 
Accretion in the poorly ionised shearing magnetic flows is dominated by the non-ideal magnetohydrodynamic (MHD) effects with the accretion flow itself being not necessarily turbulent. 
    \citet{LESUR14,BAI15,BAI17} have shown in three-dimensional MHD modelling that the Hall effect can largely control the dynamics of the laminar midplane flow yielding efficient wind accretion. 
Still, turbulent motions in MRI-inactive regions ('dead zones') can develop via hydrodynamic or gravitational instabilities. 
These require specific thermodynamic conditions, which can be used to localise their origins in discs. 
The requirement for an appropriate thermal relaxation time is a necessary but not sufficient condition for the hydrodynamic instabilities to develop: the growth rates are affected by the stresses associated with the non-ideal MHD effects. 
There are currently three identified purely hydrodynamic\footnote{we do not discuss the gravitational instability here because of the controversy over the cooling time criterion\citep{MERU11b,MERU12,PAARDEKOOPER12,HOPKINS13,BAEHR15,TAKAHASHI16}} sources of turbulent motions in protoplanetary discs: the vertical shear instability\footnote{also known as the Goldreich-Schubert-Fricke instability in the context of differentially rotating stars \citep{GOLDREICH67,FRICKE68}}\citep[][]{URPIN98,URPIN03,NELSON13,BARKER15,LIN15}, the convective overstability in its linear \citep[][]{KLAHR03a,KLAHR14,LYRA14,LATTER16} and non-linear state \citep[subcritical baroclinic instability,][]{KLAHR03a,PETERSEN07,LESUR10}, and the zombie vortex instability \citep[][]{MARCUS13,MARCUS15}. 
The first two are linear instabilities: they can drive turbulence out of an initial equilibrium. 
The zombie vortex instability is subcritical: it demands initial, finite-amplitude noise. 
The onset, linear growth, and non-linear phase of these hydrodynamic perturbations depend on thermal relaxation because of the buoyant nature of the restoring forces. 

A careful localisation of the transport mechanisms is required for modelling disc-planet interaction \citep{AIARA15}, core migration \citep{MATSUMURA07,MATSUMURA09,MORDASINI15}, and disc transport studies. 
It is further required to make use of the growing body of high-resolution, sensitive observations \citep[][]{POHL15}, which are especially expanded by the Atacama Large Millimeter Array and extreme adaptive optics systems like Spectro-Polarimetric High-contrast Exoplanet REsearch instrument (SPHERE) at the VLT. 

Furthermore, during turbulent transport, both the induced small-scale motions \citep{BRAUER08,ZSOM10} and the gas drag in local density enhancements \citep{WHIPPLE72,BARGE95,KLAHR01,KLAHR06a} can affect the dust grain growth \citep{BIRNSTIEL12} and influence the orbital migration of protoplanets \citep{NELSON04,IDA08}. 
Dust evolution in a turbulent disc with an ongoing planetesimal formation changes the dust density and its contribution to the opacity of the disc \citep{JOHANSEN06,BIRNSTIEL12}. 
The migration of protoplanets depends on the ability of the surrounding disc material to liberate the excess thermal energy, that is, the thermal relaxation \citep{LYRA10,DITTKRIST14}. 
For example, the so-called type I migration of low-mass embedded cores in locally isothermal discs leads to their fast \citep[on a timescale shorter than the disc lifetime,][]{KORYCANSKY93} depletion. 
A finite relaxation rate, however, can lead to a reversal of the migration direction \citep{PAARDEKOOPER06,KLEY08,KLEY09} by enhancing the corotation torque \citep{BARUTEAU08}. 

The computational challenge of incorporating radiative transfer into dynamical modelling as well as the uncertainties in opacities and their secular change precludes the accurate treatment of energy evolution in accretion discs. 
A sensible parametrisation for numerical studies is the thermal relaxation timescale in the energy evolution equation. 
3D hydrodynamic simulations with full radiative transport necessitate a realistic a priori estimate of the relaxation timescale as a function of disc parameters. 
Both the set-up of the numerical experiments and their physical interpretation require effective thermal relaxation timescale. 

In this paper, we present a simple method to estimate the relaxation timescale of linear temperature perturbations (Sect.~\ref{sec2:trt}). 
The formulae were derived from non-compressive linear analysis of the energy evolution equation. 
Given the mean opacities and a predefined disc structure (Sect.~\ref{sec3:setup}), the technique allows mapping of the radiative relaxation timescale over the disc interiors (Sect.~\ref{sec4:res}). 
Such maps directly constrain the origins of the hydrodynamic turbulence. 
We discuss the restrictions of the method and measures for further improvement in Sect.~\ref{sec5:disc}. 
A brief summary in Sect.~\ref{sec6:summ} closes the paper. 

\section{Radiative thermal relaxation}\label{sec2:trt}
Thermal relaxation timescale is a characteristic evolution timescale of linear temperature perturbations. 
In such dilute media as protoplanetary discs (PPDs), the relaxation is chiefly supplied by means of radiative transport. 
It is convenient to express the thermal relaxation in terms of a dimensionless parameter $\beta = t / \Omega^{-1}$, where $\Omega^{-1}$ is the sound crossing time over one pressure scale height $H$. 

To isolate the characteristic timescale of the relaxation, we have considered the evolution of only the temperature (actually, thermal and radiation energy densities, see Appendix~\ref{appA:EoT}). 
We neglected the associated changes in pressure and velocity field \citep[cf.][]{LYRA14}.
Linearising the energy evolution equation, we find the characteristic decay timescale of a Fourier mode $\delta T_k = (T - T_\mathrm{0})_k$:
\begin{equation}
    t_{\mathrm{relax},k} = \frac{ \left| \delta T_k \right| }{ \delta\dot{T}_k }.
    \label{eq:trelax0}
\end{equation}
Here $T$ is the perturbed temperature, $T_\mathrm{0}$ the equilibrium temperature ($\dot{T}_\mathrm{0} = 0$) and $\delta \dot{T}$ designates the time derivative of the perturbation. 

To estimate the radiative relaxation timescale, we calculated the characteristic timescales of the underlying processes. 
The relevant processes differ depending on the optical thickness of the mode (the optical thickness of matter encompassed within a spatial Fourier wavelength). 
An appropriate frequency-averaged measure of this optical thickness is the Rosseland optical depth (see Sect.~\ref{sec2:tthick}). 
This distinguishes optically thick and optically thin relaxation regimes with characteristic timescales $t_\mathrm{thick}$ and $t_\mathrm{thin}$, respectively. 
Because the underlying processes operate simultaneously, one can approximate the relaxation timescale by the longest timescale amongst the two \citep[cf.][]{HUBENY90}:
\begin{equation}
    t_\mathrm{relax} = \max\left(t_\mathrm{thick}, t_\mathrm{thin}\right). 
    \label{eq:trelax}
\end{equation}
Below we introduce each of these timescales. 

    \subsection{Optically thin relaxation regime}\label{sec2:tthin}
The thermal energy of the gas in a PPD dominates the thermal energy of the dust due to the high gas-to-dust mass fraction: 
\begin{equation}
    \frac{ \rho_\mathrm{gas} }{ \rho_\mathrm{dust} } = \eta^{-1} \sim 10^{2}. 
    \label{eq:eta}
\end{equation}
However, the dust emissivity, as measured by the Planck mean opacity, is comparable or exceeds that of the gas at low temperatures. 
The radiative efficiency of the gas with the specific thermal capacity $C_\mathrm{v}$ is expressible in terms of the thermal emission timescale (see Appendix~\ref{appB:themti})
\begin{equation}
    t_\mathrm{emit} = \frac{ C_\mathrm{v} }{ 16\max\left( \kappa^\mathrm{d}_\mathrm{P}, \kappa^\mathrm{g}_\mathrm{P} \right)\sigma T^{3} }
    \label{eq:temit}
\end{equation}
with $\sigma$ being the Stefan-Boltzmann constant, $\kappa^\mathrm{d/g}_\mathrm{P}$ the Planck mean opacity of the dust (per unit gas mass) or the gas, respectively. 
Timescale \eqref{eq:temit} is evaluated locally, and, therefore, does not depend on perturbation wavelength. 
This is a sensible approximation for optically thin modes. 
The optically thin emission timescale as given by Eq.~\eqref{eq:temit} is the shortest achievable relaxation timescale for the given material in local thermodynamic equilibrium (LTE). 

At dust temperatures in the range $650 < T < 1\,300\mbox{ K}$, when the abundant volatile materials are already evaporated, the values of the Planck mean opacities of the dust and the gas are comparable \citep{MALYGIN14,SEMENOV03}. 
This means that the gas dominates the thermal emission. 
Here, we have approximated the medium as being made of three sorts of species: first -- the most abundant poorly emitting gas (thermal carriers: H$_2$, He); second -- the remaining gas (including emitters, like H$_2$O, TiO, SiO, CO, etc.), and third -- the dust. 
Collision with electrons are taken into account in the equilibrium radiative processes via the Planck opacities.  
The bulk of thermal energy of the gas is stored in the kinetic energy of the most abundant H$_2$ molecules, which have no permanent dipole moment\footnote{The less abundant HD isotopologue, though, has a permanent dipole moment that permits its detection \citep{BERGIN13}} and can transfer their energy to the species with higher emissivity (both dust and gas, see Appendix~\ref{appB:coll}) via collisions. 
The speed of this coupling was measured by respective collision timescale
\begin{equation}
    t_\mathrm{coll} = \frac{1}{n\sigma v}. 
    \label{eq:tcoll}
\end{equation}
In a linear approximation (small temperature perturbations) thermal capacity of neither the dust nor the emitting gas is exhausted. 
Thus, the gas and the dust emitters each set a characteristic relaxation timescale
\begin{equation}
    t^\mathrm{d/g}_\mathrm{relax} = \max\left( t^\mathrm{d/g}_\mathrm{emit}, t^\mathrm{d/g}_\mathrm{coll} \right).
    \label{eq:trelaxdg}
\end{equation}
If $t^\mathrm{d}_\mathrm{coll} > t^\mathrm{d}_\mathrm{emit}$ (or $t^\mathrm{g}_\mathrm{coll} > t^\mathrm{g}_\mathrm{emit}$), we say that the dust (or gas) is collisionally decoupled. 
Relaxation via dust and gas channels operates simultaneously, the fastest channel determining the optically thin relaxation of the whole material: 
\begin{equation}
    t_\mathrm{thin} = \min\left( t^\mathrm{d}_\mathrm{relax}, t^\mathrm{g}_\mathrm{relax} \right). 
    \label{eq:tthin}
\end{equation}
In the so-called molecular layers above the midplane, one has to account for photoelectric heating, mutual irradiation of the dust and the gas, photochemistry, etc. \citep{KAMP01,AKIMKIN13,HENNING13}. 
Those are not taken into account in this study, where we have focussed on the bulk amount of disc gas in the midplane. 

    \subsection{Optically thick relaxation regime}\label{sec2:tthick}
A perturbation with wavelength exceeding the diffusion length scale 
\begin{equation}
    l_\mathrm{diff} = \frac{1}{\kappa_\mathrm{R}\rho}
    \label{eq:ldiff}
,\end{equation}
can not relax as fast as the local thermal emission timescale because the thermal photons get absorbed and re-emitted multiple times before escaping the perturbed medium. 
Relaxation timescale of mode $\lambda$ can be approximated by the corresponding diffusion timescale (see Appendix~\ref{app:A})
\begin{equation}
    t_\mathrm{thick} = \frac{1}{\tilde{ D } k^{2}} \propto \tau_\mathrm{R} \lambda \rho T^{-3} , 
    \label{eq:tthick}
\end{equation}
with the effective opacity in the frequency-averaged diffusion coefficient $\tilde{D}\propto(\kappa\rho)^{-1}$ being the Rosseland opacity $\kappa_\mathrm{R}$ and
\begin{equation}
    \tau_\mathrm{R} = \int_{-\lambda/2}^{+\lambda/2}\kappa_\mathrm{R}\rho\,dl.
    \label{eq:tauR}
\end{equation} 
being the Rosseland thickness over $\lambda$. 
One expects the optically thick timescale \eqref{eq:tthick} to approximate the relaxation timescale~\eqref{eq:trelax}, if the medium encompassed within the perturbation wavelength is optically thick, $\tau_\mathrm{R}>1$ (see Sect.~\ref{sec4:rr} for a refinement). 

\section{Set-up}\label{sec3:setup}
    \subsection{Density model}\label{sec3:density}
We have adopted an axisymmetrical disc density structure from \citet{DZYURKEVICH13} to allow comparison with the MHD studies on the magnetically dead zone. 
The central source is a solar-like star with $M_\mathrm{*} = 1 M_\odot$. 
The disc has a surface density profile
\begin{equation}
    \Sigma = \Sigma_\mathrm{c} \left( \frac{R}{R_\mathrm{c}} \right)^{-p} \exp \left[ -\left( \frac{R}{R_\mathrm{c}} \right)^{2-p} \right]
    \label{eq:sigma}
,\end{equation}
with $\Sigma_\mathrm{c}$ being the surface density at the cut-off radius $R_\mathrm{c}$: 
\begin{equation}
    \Sigma_\mathrm{c} = \left( 2 - p \right) \frac{M_\mathrm{disc}}{2\pi R_\mathrm{c}^{2}}.
    \label{eq:sigmac}
\end{equation}
The surface density is normalised to $\Sigma = 1700\mbox{ g cm}^{-2}$ at $1\mbox{ au}$. 
Index $p$ of $0{.}9$ and the cut-off radius of $40\mbox{ au}$ is chosen following \citet{ANDREWS09} to match the minimum-mass solar nebula at $1$ and $100\mbox{ au}$. 
$M_\mathrm{disc}$ is the total mass of the disc. 
The volumetric density is readily unfolded for a vertically isothermal case 
\begin{equation}
    \rho = \frac{\Sigma}{\sqrt{2\pi}H} \exp\left( \frac{-z^{2}}{2 H^{2}} \right),
    \label{eq:rho}
\end{equation}
where $H$ is the local pressure scale height. 

    \subsection{Temperature model}\label{sec3:temperature}
We adopt a vertically isothermal radially stratified temperature profile following \citet{DZYURKEVICH13}
\begin{equation}
    T = T_\mathrm{0} \left( \frac{R_\mathrm{in}}{R} \right)^{1/2},
    \label{eq:temperature}
\end{equation}
with $T_\mathrm{0} = 280\mbox{ K}$ at $R_\mathrm{in} = 1\mbox{ au}$. 
Since we focus on the bulk disc matter in the midplane region, we ignore upper disc layers and hence vertical $T$-gradient. 
A lower limit of $10\mbox{ K}$ stops the power law decrease at $\sim250\mbox{ au}$. 
This is a typical lowest temperature in outer midplane of the discs around Sun-like stars predicted by detailed radiative transfer models, such as RADMC-3D \citep{DULLEMOND12}. 

\begin{figure}[htpb]
    \resizebox{\hsize}{!}{\includegraphics{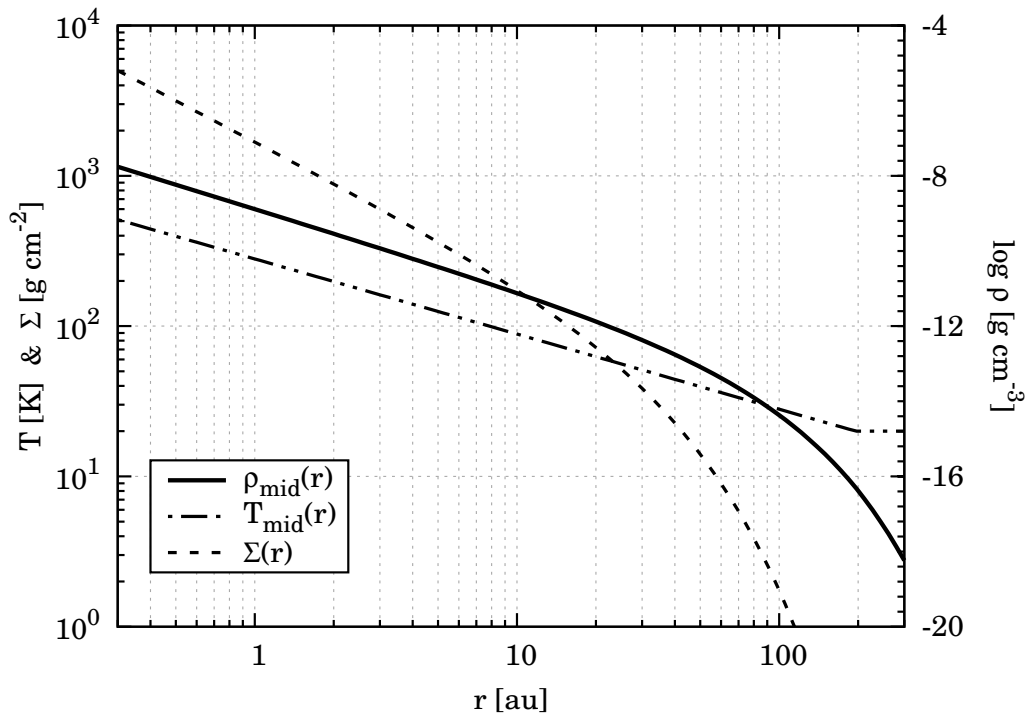}}
    \caption{Disc structure as adopted from \citet{DZYURKEVICH13}. 
    Displayed are the surface density profile (dashed): power law with exponential roll-off; 
    the temperature power law profile bounded from below (dot-dot-dashed);
    the corresponding volumetric density through the midplane (solid, the values at the right ordinate). 
}
    \label{fig:disc_structure}
\end{figure}

The profiles of surface density, volumetric density, and the midplane temperature are given in Fig.~\ref{fig:disc_structure}. 
The highest value of $M_\mathrm{disc}=0{.}1M_\odot$ considered for this model assures the configuration is stable against self-fragmentation, according to the Toomre $Q$-criterion. 

    \subsection{Opacities}\label{sec3:opacity}
\begin{figure}[thpb]
    \resizebox{\hsize}{!}{\includegraphics{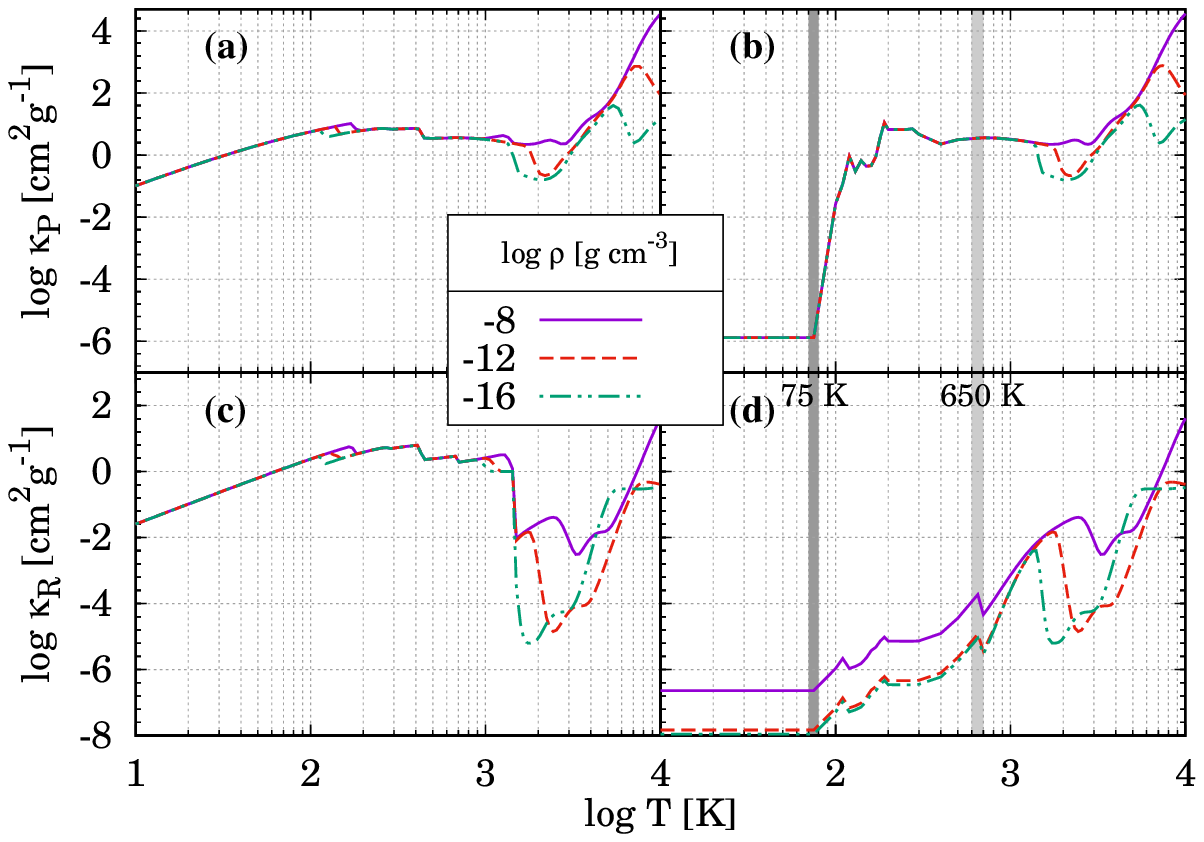}}
    \caption{Planck (\textit{top row}) and Rosseland (\textit{bottom row}) mean opacities. 
    Panels (a), (c): joint dust \citep[][homogenous IPS]{SEMENOV03} and gas \citep{MALYGIN14} opacities. 
    Panels (b), (d): pure gas opacity, joint data from \citet{FREEDMAN08} and \citet{MALYGIN14}. 
    Data from \citet{FREEDMAN08} is linearly scaled down in density in log space. 
    The large gas opacity model extrapolates to lower temperatures the values at $650\mbox{ K}$ \citep[the low-temperature cut of data from][]{MALYGIN14}; the small gas opacity model extrapolated to lower temperatures the values at $75\mbox{ K}$ \citep[the low-temperature cut of data from][]{FREEDMAN08}. 
    The numbers on the legend indicate the logarithm base ten of the density in cgs units. }
\label{fig:opacity}
\end{figure}

We used tabulated dust and gas mean opacities (Fig.~\ref{fig:opacity}). 
The dust opacity is that of iron-poor homogeneous silicate (IPS) spheres from \citet{SEMENOV03}, the gas opacity is that of the solar-mixture gas from \citet{MALYGIN14}. 
The total opacity is the direct sum of the gas and the dust component. 
Figures~\ref{fig:opacity}a,c display the combined data. 
Gas chemistry and opacity in outer disc regions are poorly constrained observationally \citep[see][and references therein]{DUTREY14}. 
In calculations with significant dust depletion, we considered two cases of pure gas opacity: (i) data from \citet{MALYGIN14} with the $650\mbox{ K}$ cut-off at low temperatures; (ii) data from \citet{MALYGIN14} complemented with scaled-down (in density) atmospheric opacities from \citet{FREEDMAN08} with the lower boundary of $75\mbox{ K}$. 
For yet lower temperatures, the value at the lowest available temperature is always used. 
Data from \citet{FREEDMAN08} accounts for the settling of condensates in a gravitational field (e.g. in cool dwarves) but neither \citet{FREEDMAN08} nor \citet{MALYGIN14} account for the freeze out on grain surfaces. 
However, the data from \citet{SEMENOV03} takes into account freeze out of volatiles such as organics, H$_2$O, and CO, the main absorbers at cold temperatures in the gas phase. 

Planck opacities from \citet{FREEDMAN08} have relatively high values in the interval $180 - 320\mbox{ K}$ and a steep decline below $\approx120\mbox{ K}$ (Fig.~\ref{fig:opacity}b). 
Rosseland opacities from \citet{FREEDMAN08} exceed those from \citet{MALYGIN14} in the interval $480 - 700\mbox{ K}$ and then drop by $\sim3$ orders of magnitude at $75\mbox{ K}$ (Fig.~\ref{fig:opacity}d). 
We will refer to the data complemented with \citet{FREEDMAN08} as the small gas opacity case and to the data extrapolated from \citet{MALYGIN14} as the large gas opacity case. 
The two models can be treated as rough borders of the poorly-constrained parameter space. 
One of the motivations beyond the large gas opacity case can be an increase in the gas-phase opacity (e.g. a release of molecules not taken into account in \citet{FREEDMAN08}) after local dust evaporation. 
When varying the dust-to-gas mass ratio (Sect.~\ref{sec4:dtgmr}), we modified the dust opacity by
\begin{equation}
    \kappa^\mathrm{d} = \frac{\eta'}{\eta}\kappa^\mathrm{d}, 
    \label{eq:kappareddust}
\end{equation}
where $\eta'$ is the new dust-to-gas mass ratio and $\eta=0{.}014$ the nominal dust-to-gas mass ratio, for which the dust opacity tables were calculated \citep[see Table 1 in][ case of IPS]{SEMENOV03}.
The gas opacity stays unchanged. 

\section{Results}\label{sec4:res}
In this section, the values for the relaxation rate are dimensionless numbers $\Omega t$, with $\Omega$ being the local Keplerian frequency. 
The radiative diffusion timescale is calculated for diffusion in the vertical direction unless explicitly stated otherwise. 
The spatial perturbation wavelength is conveniently measured in local pressure scale heights: $\Tilde\lambda = \lambda / H$, or $ \Tilde k = 2\pi\Tilde\lambda^{-1} = kH$. 

    \subsection{Relaxation regimes: verification}\label{sec4:rr}
One expects the transition from the optically thin to the optically thick relaxation to happen when the perturbation wavelength $\lambda$ becomes of the order the diffusion length scale $l_\mathrm{diff}$. 
Equating \eqref{eq:temit} and \eqref{eq:tthick} while letting $\lambda = l_\mathrm{diff}$ gives
\begin{equation}
    \int_{-\lambda/2}^{+\lambda/2}\kappa_\mathrm{R}\rho\,dl = \tau_\mathrm{R} = \frac{\left( 2\pi \right)^{2}}{3}\frac{\langle\kappa_\mathrm{R}\rangle}{\kappa_\mathrm{P}},
    \label{eq:tauRtrans}
\end{equation}
where $\langle\kappa_\mathrm{R}\rangle$ is the spatial average of the Rosseland opacity over $\lambda$. 
A relevant tracer of the transition $t_\mathrm{thin} = t_\mathrm{thick}$ is the Rosseland optical depth over the region of the same sign of the perturbation amplitude (between two subsequent nodes)\footnote{For dust continuum, $\tau_\mathrm{R}\approx1$ at the transition but for pure gas this value is lower due to the larger difference between the Planck and the Rosseland mean opacities.},
\begin{equation}
    \tau_\mathrm{R,\lambda/2} = \int_{-\lambda/4}^{+\lambda/4}\kappa_\mathrm{R}\rho\,dl.
    \label{eq:tauRhalf}
\end{equation} 

\begin{figure}[thpb]
    \centering
    \resizebox{\hsize}{!}{\includegraphics{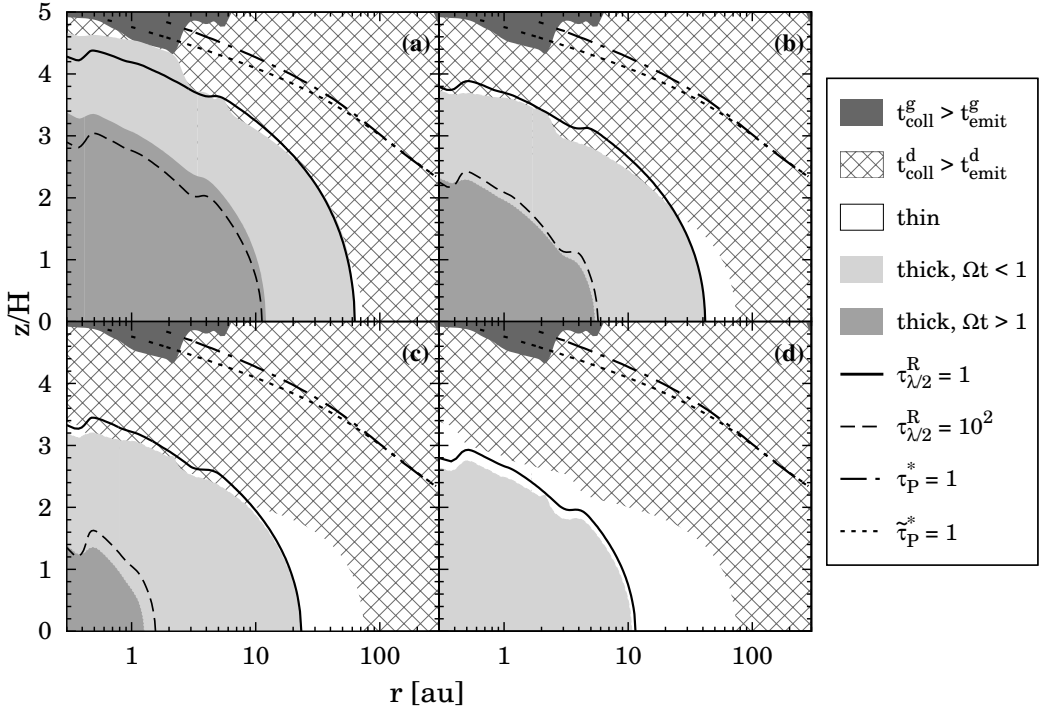}}
   \caption{ Verification of the maximum-time criterion. 
        The four panels correspond to different perturbation wavelengths, $\Tilde\lambda = 2$ (a), $0{.}5$ (b), $0{.}1$ (c), and $0{.}02$ (d). 
        The cross-hatched pattern identifies regions of the dust collisional decoupling, the dark grey areas -- of the gas collisional decoupling (see the legend); the white areas refer to the fastest radiative relaxation via thermal emission; optically thick relaxation is marked with light grey, and the grey additionally highlights the dynamically slow relaxation ($\Omega t > 1$). 
        Rosseland depth $\tau_\mathrm{R,1/2}=1$ (solid lines) approximates the transition between the optically thin and the optically thick regimes. 
        The absorption of the stellar light is shown for the two gas opacity models: the small gas opacity (dot-dashed) with single-temperature Planck means from \citet{FREEDMAN08}   and the large gas opacity (dotted) with two-temperature Planck means from \citet{MALYGIN14}. }
    \label{fig:verif}
\end{figure}

Figure~\ref{fig:verif} presents a verification of the maximum time criterion introduced in Sect.~\ref{sec2:trt}. 
In all panels of Fig.~\ref{fig:verif} we have marked $\tau_\mathrm{R,\lambda/2}=1$, which closely (but not exactly) traces the transition between the optically thick and the optically thin relaxation \eqref{eq:trelax}. 
In an opaque inner disc, even the high-frequency modes are optically thick and relax on the radiative diffusion timescale. 
This can be either dynamically slow, $\Omega t > 1$ (up to $\Omega t\sim10^{3}$), or dynamically fast, $\Omega t < 1$ (down to $\Omega t\sim10^{-3}$). 
The fastest achievable radiative relaxation, $\Omega t \sim 10^{-4} - 10^{-3}$, is that of the optically thin modes in the yet coupled regions ($t_\mathrm{coll} < t_\mathrm{emit}$, see Fig.~\ref{fig:verif}). 

Absorption of the stellar photons ($T_*=5777\mbox{ K}$) is shown in Fig.~\ref{fig:verif} for the two gas opacity models. 
In the small gas opacity case, the single-temperature Planck means from \citet{FREEDMAN08} are used to calculate the optical depth:
\begin{equation}
    \tau^*_\mathrm{P} = \int \left( \kappa^\mathrm{g}_\mathrm{P}\left( T_* \right) + \kappa^\mathrm{d}_\mathrm{P} \left( T_* \right) \right) \rho dR = \int \kappa^\mathrm{g}_\mathrm{P}\left( T_* \right)\rho dR . 
    \label{eq:taupsmall}
\end{equation}
In the large gas opacity case, the two-temperature\footnote{The temperature of the radiation field $B_\nu\left( T_* \right)$ differs from the gas temperature $T_\mathrm{g}$} Planck means are available from \citet{MALYGIN14}:
\begin{equation}
    \tilde{\tau}^*_\mathrm{P} = \int \kappa^\mathrm{g}_\mathrm{P}\left( T_*, T_\mathrm{g} \right)\rho dR
    \label{eq:tauplarge}
.\end{equation}
In both cases, the integration is done along the ray of constant polar angle in spherical coordinates between $0{.}3\mbox{ au}$ and $300\mbox{ au}$. 
The contribution from innermost regions ($<0{.}3\mbox{ au}$) is neglected. 

\begin{figure}[!tb]
    \resizebox{\hsize}{!}{\includegraphics{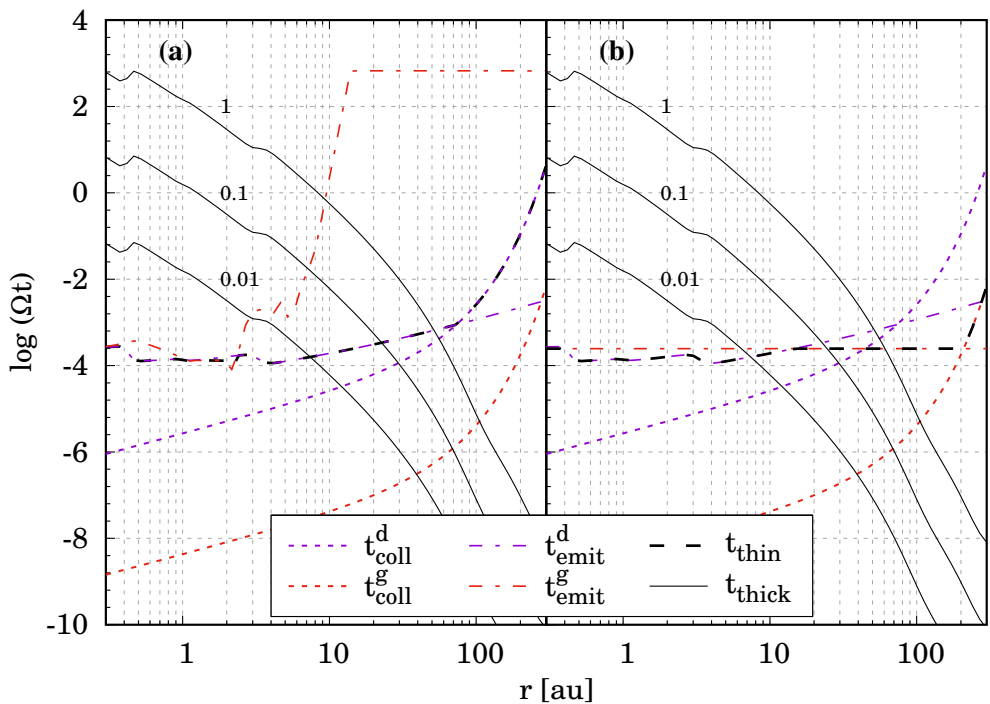}}
   \caption{ Timescales involved in thermal relaxation. 
       Values are scaled with local Keplerian frequency $\Omega$. 
       Panel (a): the small gas opacity model, panel (b): the large gas opacity model. 
       The collision timescales are shown with the short dashed lines (grey for gas-to-gas, black for dust-to-gas); the LTE emission timescales are shown with the dot-dashed lines (grey for gas, black for dust); the resulting timescale of the optically thin relaxation is shown with the long dashed lines. 
       The solid lines show the timescale of radiative diffusion in the vertical direction from the midplane for perturbation wavelengths $\Tilde\lambda = 1, 10^{-1}, 10^{-2}$ (top to bottom). 
    \label{fig:ct}}
\end{figure}

Figure~\ref{fig:ct} displays a comparison between LTE emission timescale \eqref{eq:temit}, collision timescale \eqref{eq:tcoll}, and radiative diffusion timescale \eqref{eq:tthick} for midplane modes and the fiducial dust-to-gas mass ratio of $0{.}014$. 
The two panels correspond to the two gas opacity models. 
The optically thick relaxation, being dominated by the dust, is not sensitive to the gas opacity model. 
Relaxation timescale in this regime drops with radius, its absolute value depends on the perturbation wavelength. 
Low-frequency modes $\Tilde\lambda=1$ relax dynamically slow ($1 < \Omega t < 10^{3}$) within first $\approx8\mbox{ au}$. 
The farther out the faster they relax: the relaxation timescale drops to $\approx10^{-3}\Omega^{-1}$ at $\approx50\mbox{ au}$. 
High-frequency yet optically thick modes $\Tilde\lambda\leq10^{-2}$ always relax dynamically fast ($\Omega t < 1$). 
Optically thick relaxation timescale \eqref{eq:tthick} significantly underestimates the relaxation timescale if a mode becomes optically thin (the long-dashed lines in Fig.~\ref{fig:ct}).  
This is an artefact of the free-streaming limit in the flux-limited diffusion approximation (see Sect.~\ref{sec5:disc}). 

The gas opacity model becomes important for optically thin modes. 
In the large gas opacity case (Fig.~\ref{fig:ct}b), the optically thin midplane modes relax on the gas thermal emission timescale, $\Omega t\approx2{.}5\times10^{-4}$, from $15\mbox{ au}$ to $200\mbox{ au}$ in the midplane.

\begin{figure}[!tb]
    \resizebox{\hsize}{!}{\includegraphics{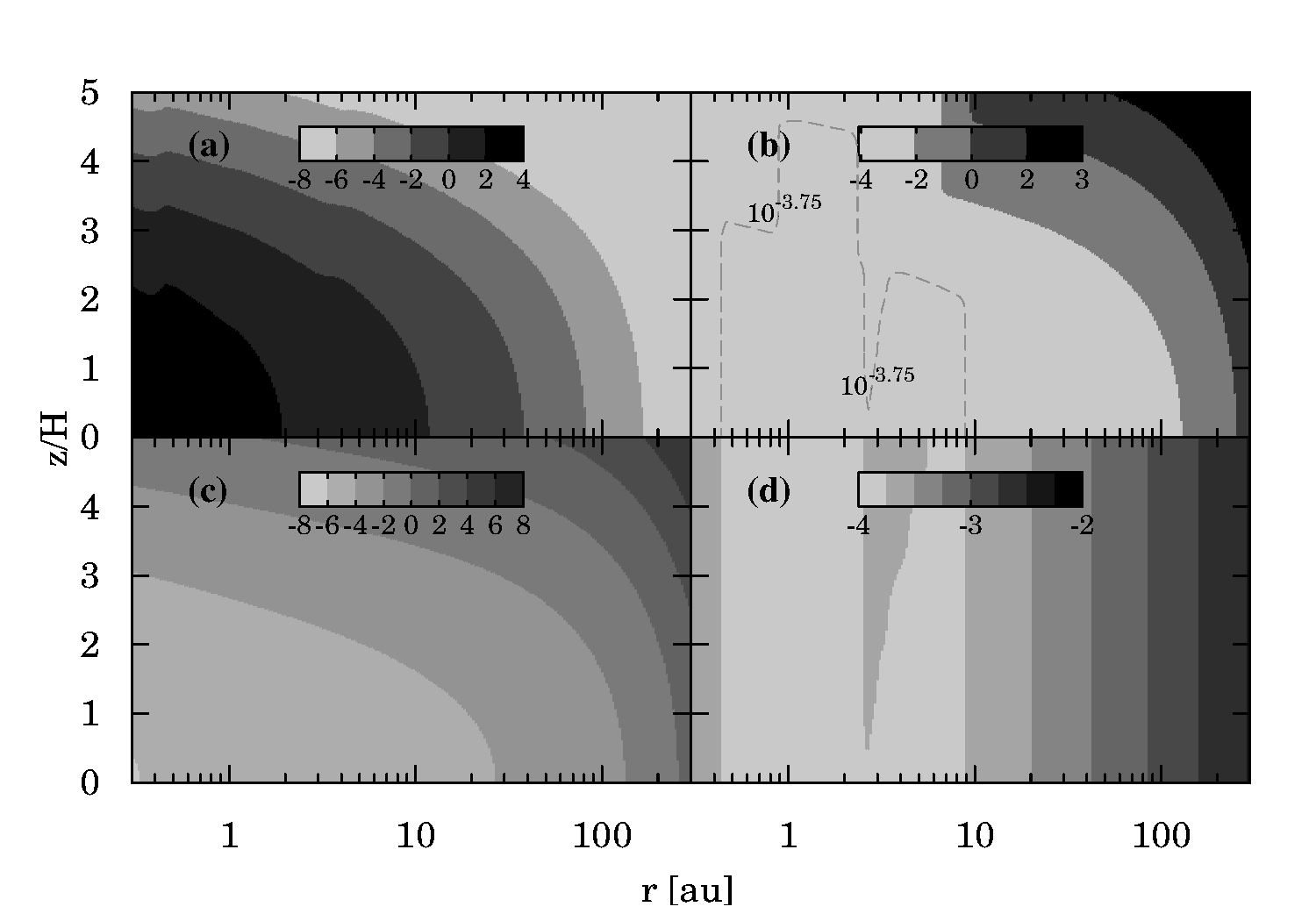}}
    \caption{ Maps of optically thick (a) and optically thin (b) relaxation timescales. 
        Vertical perturbations with $\Tilde\lambda_z = 2$. 
        Plotted in colour are the values of $\log\left(\Omega t\right)$. 
        The two bottom panels show the dust collision time (c), and the dust LTE emission time (d).
        \label{fig:irtm}}
\end{figure}

Figure~\ref{fig:irtm} maps the timescales of the processes that contribute to the thermal relaxation all over the disc. 
The map of the radiative diffusion timescale (Fig.~\ref{fig:irtm}a) is plotted for $\Tilde\lambda = 2$. 
Figure~\ref{fig:irtm}b maps the optically thin relaxation timescale. 
This is a combination of the collision and the thermal emission timescales according to \eqref{eq:tthin}. 
Figures~\ref{fig:irtm}c and d also show the dust-to-gas collision timescale and the dust LTE emission timescale, respectively. 
The assumed vertically unstratified temperature structure is captured in Fig.~\ref{fig:irtm}d. 
The size distribution of the grains (see Appendix~\ref{appB:coll}) as well as the dust-to-gas mass ratio are uniform over the disc. 
A wedge-like structure at around $3\mbox{ au}$ in Figs.~\ref{fig:irtm}b,d is due to the water ice line.
    \subsection{Perturbation wavelength}\label{sec4:pw}
At a given moment of time, the density, temperature, and opacities are defined at any point in the disc. 
Then the regime of radiative relaxation -- optically thick or thin -- is decided by the perturbation wavelength. 

\begin{figure}[!tb]
    \resizebox{\hsize}{!}{\includegraphics[width=17cm]{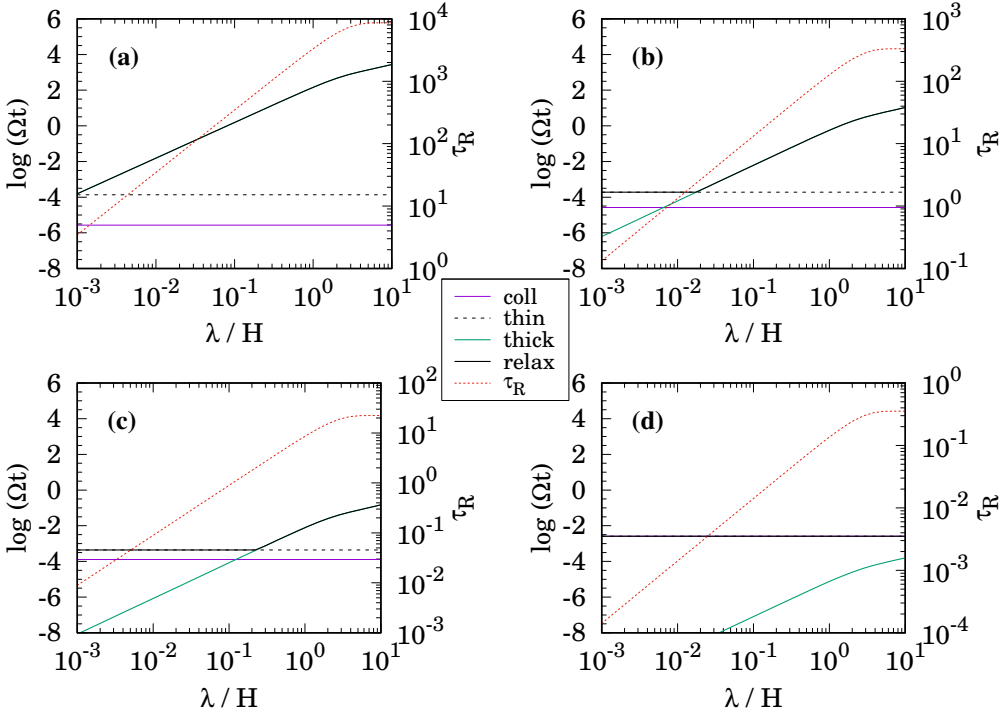}}
    \caption{Relaxation timescale versus perturbation wavelength of midplane modes at different radial separations from the central star:
          $1\mbox{ au}$ \textbf{(a)}, 
         $10\mbox{ au}$ \textbf{(b)}, 
         $30\mbox{ au}$ \textbf{(c)}, 
        and $100\mbox{ au}$ \textbf{(d)}. 
        In each panel: the relaxation timescale (thick solid), the optically thin timescale (short dashed), the optically thick timescale (long dashed), the dust-to-gas collision timescale (dot-dashed), the Rosseland thickness of the mode (dotted, the values on the right ordinate). 
}
    \label{fig:rtvw}
\end{figure}

Figure~\ref{fig:rtvw} shows relaxation timescale of vertical modes as a function of their wavelength at different radial locations. 
Relaxation via thermal emission puts a wavelength-independent lower limit on relaxation timescale (indicated in Fig.~\ref{fig:rtvw}). 
In the optically thick regime, the dependence of the relaxation timescale on $\lambda$ changes itself with $\lambda$: for $\Tilde\lambda\la1$,
\begin{equation}
    t_\mathrm{thick}\propto\tau_\mathrm{R}\lambda\propto\lambda^{2}, 
    \label{eq:lamsqr}
\end{equation}
and for $\Tilde\lambda>1$, 
\begin{equation}
    t_\mathrm{thick}\propto\lambda
    \label{eq:lamlin}
,\end{equation}
because the Rosseland depth saturates (see the dotted lines in Fig.~\ref{fig:rtvw}). 

In the bulk of the disc, the relaxation of optically thin modes via thermal emission is dynamically fast ($\Omega t\ll1$, see Figs.~\ref{fig:rtvw}b-d), optically thick modes can be both isothermal ($\Omega t \ll 1$) and adiabatic ($\Omega t \gg 1$). 
High-frequency modes $\Tilde\lambda \ll 10^{-1}$ remain isothermal even if they are optically thick (Fig.~\ref{fig:rtvw}a, Fig.~\ref{fig:ct}). 
A transition between the two relaxation regimes occurs at larger radii for larger $\Tilde\lambda$ such that the Rosseland depth of the mode $\tau_\mathrm{R}\approx2$ at the transition (Figs.~\ref{fig:rtvw}b,c, see also Sect.~\ref{sec4:rr}). 

    \subsection{Anisotropic relaxation}\label{sec4:rvm}
\begin{figure}[thb]
    \resizebox{\hsize}{!}{\includegraphics{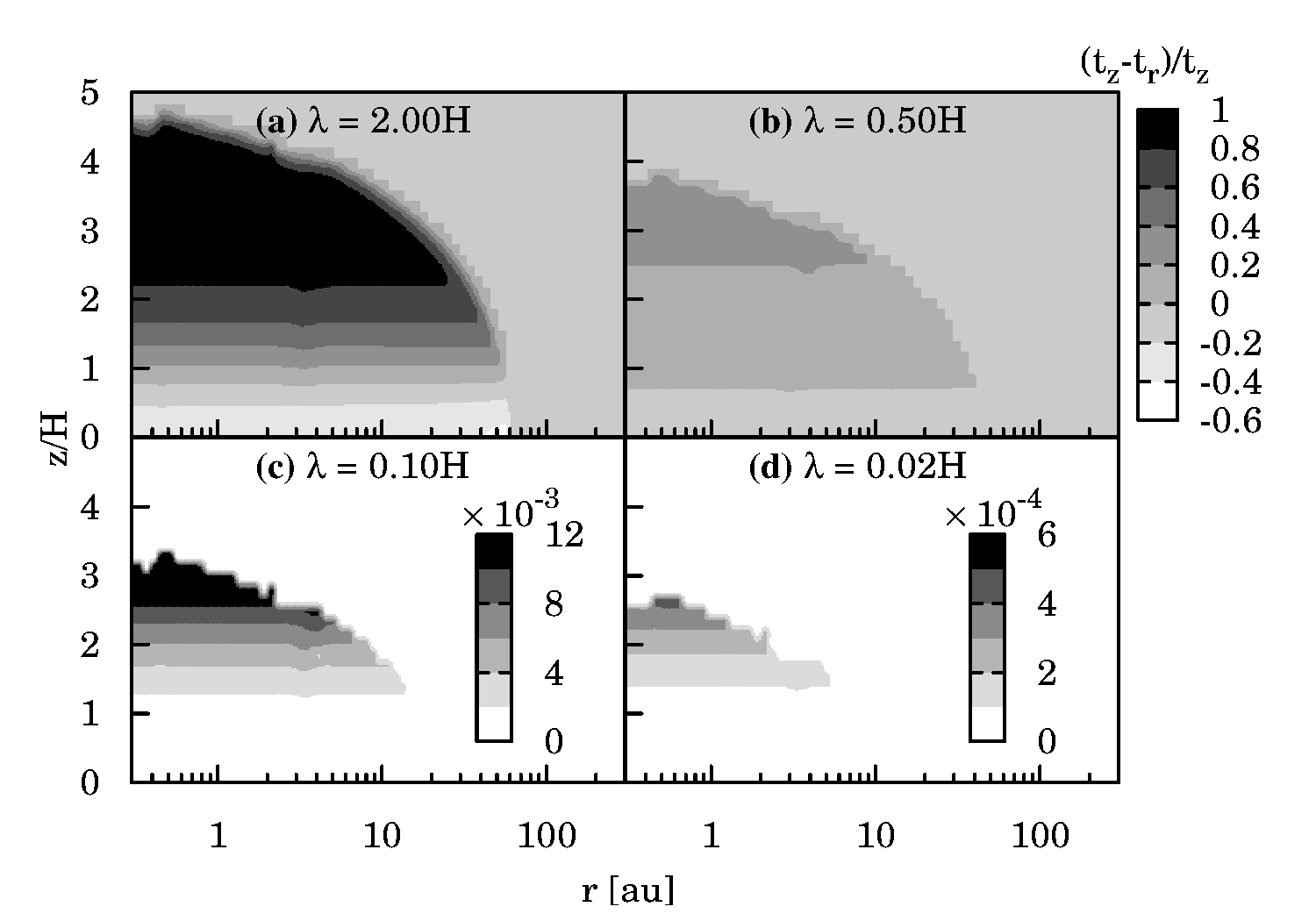}}
   \caption{Relative difference between vertical and radial relaxation timescales, $(t_\mathrm{z} - t_\mathrm{r})/t_\mathrm{z}$. 
       Each panel corresponds to a single wavelength $\lambda_z = \lambda_r$ as labelled. 
       The box on the top right sets the colour scale for the two top panels, for the two bottom panels the colour scale is inside each.  
    \label{fig:rdmaps}}
\end{figure}

Optically thick relaxation in non-isotropic media can be anisotropic itself: diffusion timescale in different directions is, in general, different. 
Figure~\ref{fig:rdmaps} maps a relative difference between relaxation timescales of vertical and radial modes of the same wavelength (anisotropy, hereafter). 
Vertical diffusion is up to $60\%$ faster than the radial diffusion for $\Tilde\lambda=2$ modes in the midplane (Fig.~\ref{fig:rdmaps}a). 
This is opposite above the midplane: diffusion in the radial direction is faster than in the vertical, and the difference is larger. 
The anisotropy is the largest when the radial direction becomes optically thin while the vertical direction remains optically thick (due to the exponential increase of the density towards the midplane). 
This is typically at $2-3H$ above the midplane. 
The maximum anisotropy is $\approx100\%$ for $\Tilde\lambda = 2$, $\approx1\%$ for $\Tilde\lambda=0{.}1$ and $\approx0{.}5\%$ for $\Tilde\lambda=0{.}02$. 
    \subsection{Dust-to-gas mass ratio}\label{sec4:dtgmr}
Models of \citet{BIRNSTIEL12} suggest that the dust growth during planet formation leads to a significant dust opacity decrease. 
In this section, we show how the radiative relaxation time changes with dust depletion degree. 
We homogeneously reduce the dust-to-gas mass ratio while leaving both the effective collision cross section per particle (the size distribution) and the dust opacity per unit dust mass (composition) unchanged for simplicity. 
Dust thermal radiation timescale \eqref{eq:temit} and dust-to-gas collision timescale \eqref{eq:tcoll} both increase with dust reduction. 
This is also expected to be so due to depletion of small grains caused by the dust growth. 

Under typical PPD conditions, the Rosseland mean is dominated by the dust continuum below some $1\,500\mbox{ K}$, but the Planck means of the gas and the dust compare down to at least $650\mbox{ K}$ owing to molecular ro-vibrational bands \citep{MALYGIN14}. 
The optically thick relaxation is determined by Rosseland opacity and hence is sensitive to the dust evolution effects. 
The gas opacity is negligible in the optically thick relaxation up until significant ($\times10^{-2}$) dust depletion. 
The optically thin regime is, in contrast, sensitive to the gas opacity model, and already at the fiducial mass ratio (Fig.~\ref{fig:ct}). 

\begin{figure}[tpbh]
    \resizebox{\hsize}{!}{\includegraphics{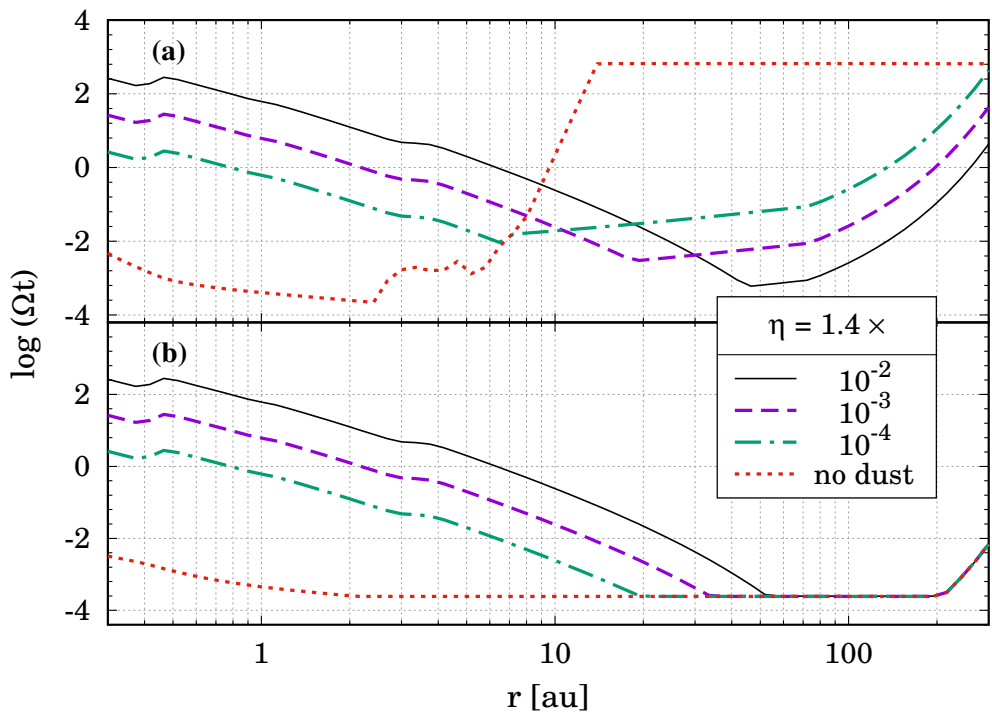}}
   \caption{ Effect of reducing dust-to-gas mass ratio $\eta$. 
       The relaxation time of midplane modes with $\lambda_r/H = 0{.}628$ versus radial location for the small gas opacity (a) and the large gas opacity (b). 
    \label{fig:reddust}}
\end{figure}

The optically thick relaxation in Fig.~\ref{fig:reddust} is calculated for radial modes $\Tilde\lambda_r = 0{.}628$ ($\Tilde k_r = 10$, with an application for the VSI, see Sect.~\ref{sec4:vsi}), the qualitative trend is similar for other wavelengths. 
Dust amount reduction yields speed-up in optically thick relaxation. 
With fewer absorbers, the Rosseland optical depth is smaller, hence, the radiative diffusion timescale is shorter. 
The relaxation timescale drops with dust reduction until it equals the optically thin relaxation timescale. 

In what way the optically thin relaxation changes with dust depletion, depends on how the Planck means of the dust and the gas compare. 
In the small gas opacity case, the optically thin relaxation slows down with the dust amount decrease (Fig.~\ref{fig:reddust}a). 
This is due to the shrinkage of the net dust emitting surface. 
In the case of zero dust (the dotted line in Fig.~\ref{fig:reddust}a), the disc is optically thin in the vertical direction and the relaxation is restrained by the gas emissivity. 
In the large gas opacity case, however, optically thin modes do not ``feel'' the dust depletion. 
This is because the gas determines the relaxation as a more efficient (larger Planck mean) yet collisionally faster-coupled emitter than the dust already at the fiducial density ratio of $0{.}014$. 

\begin{figure}[!tb]
    \resizebox{\hsize}{!}{\includegraphics{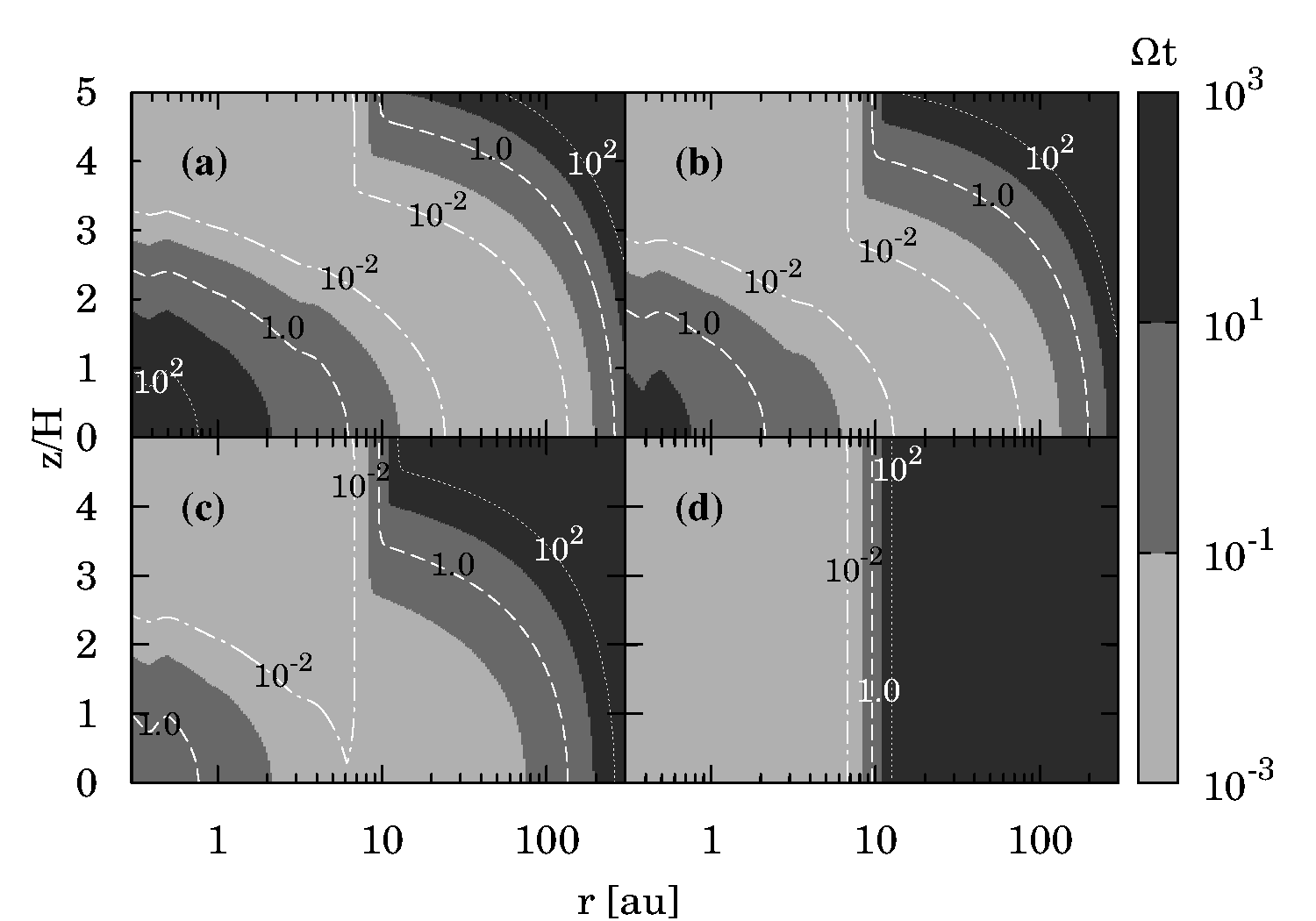}}
    \caption{ Maps of the relaxation timescale of $\Tilde k_r = 10$ perturbations for different dust depletion factors and the small gas opacity case. 
       The fiducial dust-to-gas mass ratio $\eta = 1{.}4\times10^{-2}$ (a), is then reduced by $10^{-1}$ (b), and $10^{-2}$ (c). 
       Panel (d) shows the case of zero dust. 
    \label{fig:drmap_wF}}
\end{figure}

\begin{figure}[!tb]
    \resizebox{\hsize}{!}{\includegraphics{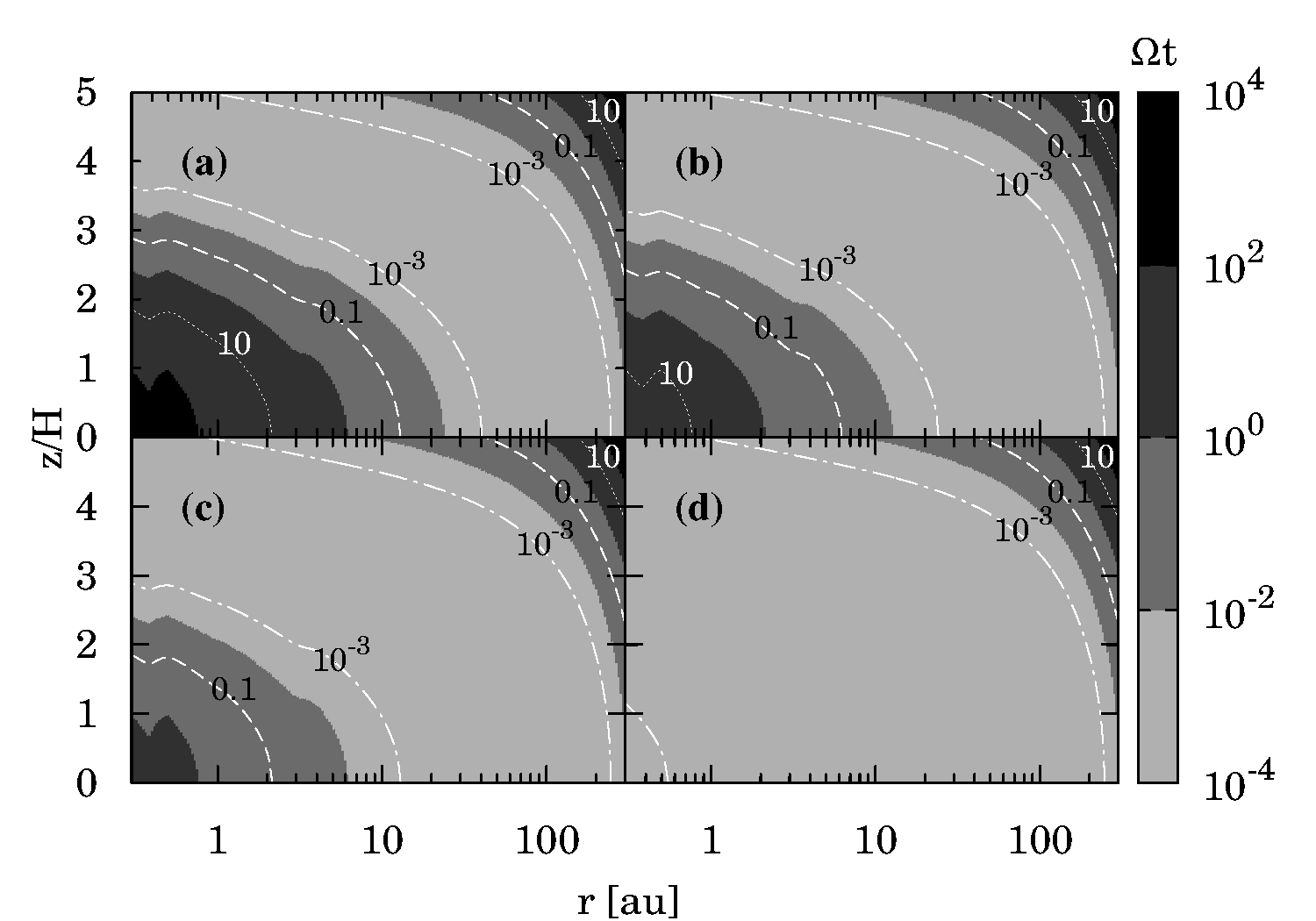}}
    \caption{ Same as Fig.~\ref{fig:drmap_wF} but for the large gas opacity case. 
    \label{fig:drmap}}
\end{figure}

Figures~\ref{fig:drmap_wF} and \ref{fig:drmap} map the relaxation timescale of $\Tilde k_r = 10$ modes (see Sect.~\ref{sec4:vsi}) under different dust depletion degrees for the small and the large gas opacity case, respectively.
The dust-to-gas mass ratio is varied evenly throughout the disc. 
Therefore, the maps do not describe the relaxation in a realistic, planet-forming environment, where the ratio can be unevenly distributed and evolve with time. 
However, these highlight the dependence of the relaxation on gas opacity and dust depletion, and indicate the necessity of taking into account the mean gas opacity in numerical modelling with dust evolution. 

    \subsection{Disc mass}\label{sec4:dm}
\begin{figure}[!tb]
    \resizebox{\hsize}{!}{\includegraphics{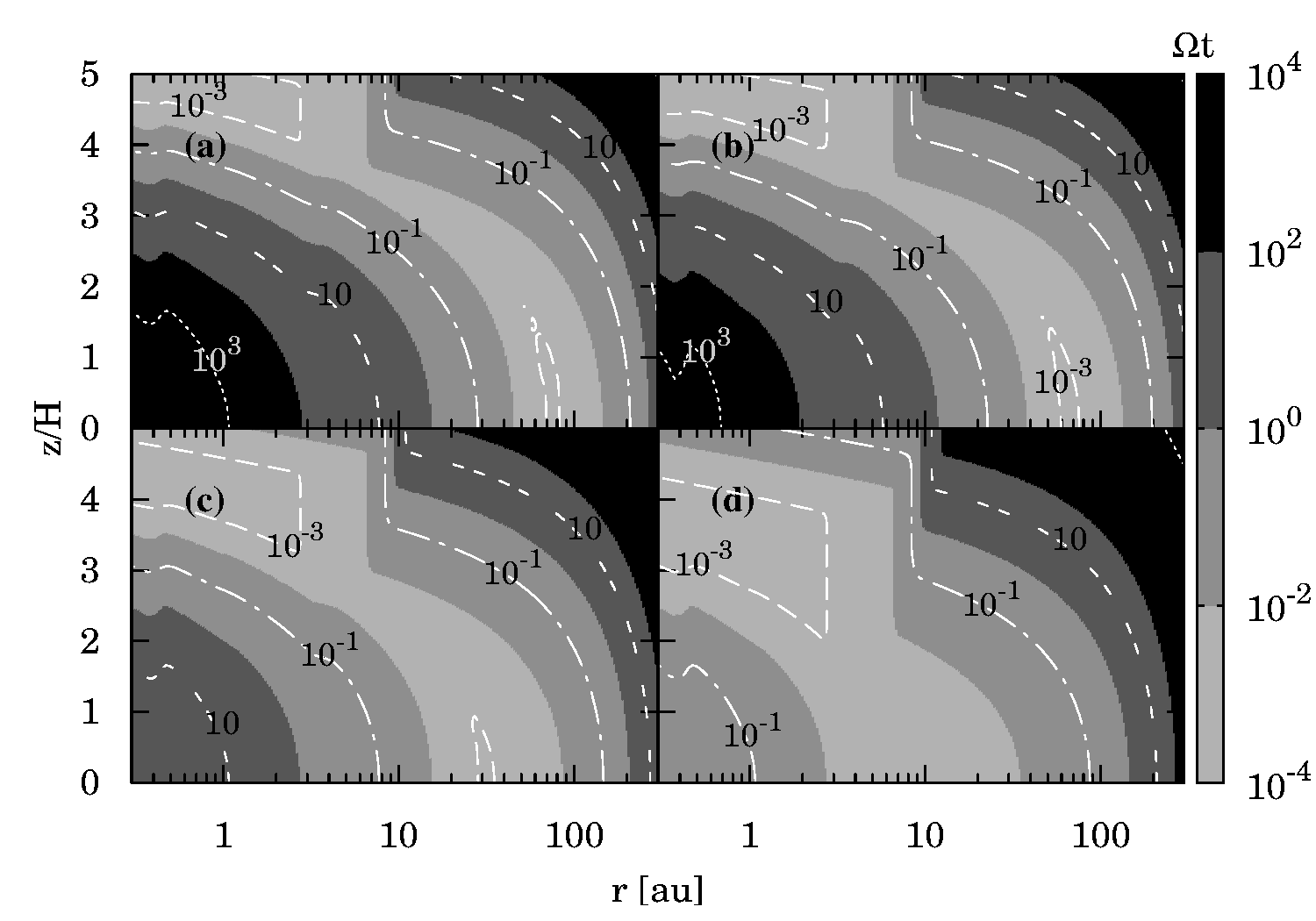}}
    \caption{ Thermal relaxation timescale of $\Tilde{\lambda} = 2$ perturbations for different disc masses: $M_\mathrm{disc} / \mathrm{M}_{\odot} = $
       $0{.}1$          \textbf{(a)}, 
       $0{.}064$        \textbf{(b)}, 
       $0{.}01$         \textbf{(c)}, 
       $0{.}001$        \textbf{(d)}. 
    \label{fig:rt4mass}}
\end{figure}

The outcome of varying disc mass is shown in relaxation maps in Fig.~\ref{fig:rt4mass}. 
The effect is due to the density change. 
The optically thick relaxation is slower in a more massive disc than in a lighter one because of the larger Rosseland depth \eqref{eq:tauRhalf} across the mode in the denser medium. 
This is responsible for the change in relaxation in the inner opaque parts of the disc: bottom left area in Figs.~\ref{fig:rt4mass}a-d. 
The collision timescales vary with density as $n^{-1}$, so the density change affects relaxation in collisionally-decoupled outer disc: top right area in Figs.~\ref{fig:rt4mass}a-d. 
The thermal emission timescale does not change with the disc mass because it depends only weakly on density (through the Planck mean opacity). 
The net result is: in more massive discs, layers of fast relaxation are narrower and located farther out from the central object, comparing to lighter discs. 

    \subsection{Hydrodynamic transport in circumstellar discs}\label{sec4:hdi}
\begin{figure}[!tb]
    \resizebox{\hsize}{!}{\includegraphics{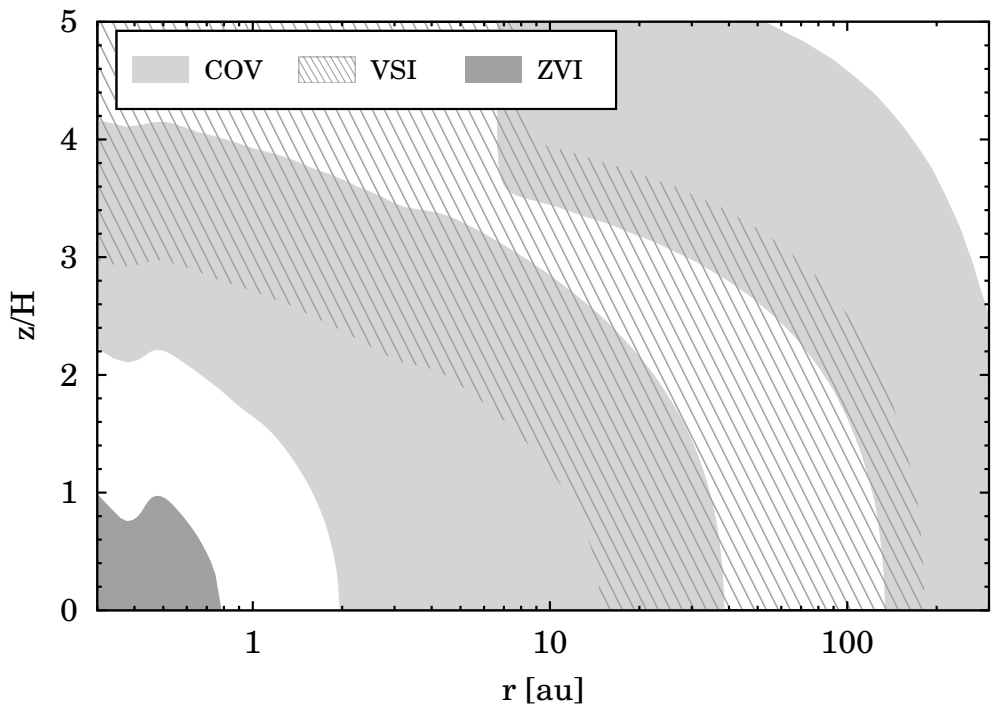}}
   \caption{ Constraints on the origin of hydrodynamic instabilities in discs. 
       ZVI (grey): $\mbox{Pe} > 10^{4}$, $\Tilde\lambda_z=2$; COV (light grey): $10^{-2} \leq \Omega t \leq 10^{2}$, $\Tilde\lambda_z=2$; VSI (hatched): $\Omega t \leq 6{.}25\times10^{-2}$, $\Tilde\lambda_r =0{.}628$. 
    \label{fig:scan}}
\end{figure}

To generate turbulence in an axisymmetric and vertically unstratified disc flow, a purely hydrodynamic mechanism must circumvent the strong centrifugal stability imposed by a positive radial specific angular momentum gradient \citep[Rayleigh's stability criterion][]{BALBUS96}. 
A general requirement for the dynamic stability of a stratified fluid, the Solberg-H\o iland conditions \citep{RUEDIGER02}, allow development of an instability for Rayleigh-stable (if unstratified) velocity fields. 
The circumvention of the stabilising radial stratification is either via vertical stratification of angular frequency $\Omega (z)$ or negative entropy gradient $\partial S/ \partial r < 0$. 
Each of the identified instabilities described below requires conditions that can be used to constrain its location within a disc (see Fig.~\ref{fig:scan}).

    \subsubsection{Vertical shear instability}\label{sec4:vsi}
The vertical shear instability (VSI) can develop in a disc with a global temperature profile, where the angular frequency is a function of both radial and vertical coordinates \citep{URPIN98,UMURHAN13}. 
The angular momentum transport by the VSI can be viewed as such a swap of the fluid elements along the curved iso-line of the specific angular momentum, which yields a state with lower total energy \citep{BARKER15}. 
The weak vertical stratification in the orbital motion ($\Omega\left( z \right)$) permits circumvention of the centrifugal stability via driving vertically elongated (and slightly tilted) but radially narrow flows. 
Vertically elongated disturbances $(k_r/k_z\gg1)$ can overcome the Coriolis stabilisation. 
However, motions of this kind are impeded by vertical buoyancy in discs with a stable vertical stratification. 
The VSI requires sufficiently fast thermal relaxation to overcome the stabilising vertical buoyancy \citep{URPIN03,LIN15}. 
The angular momentum transfer rates resulting from the VSI are modest: $\alpha\approx10^{-4}$ \citep[as found so far in numerical modelling by][]{ARLT04,NELSON13,STOLL14}. 

\citet{NELSON13} performed a detailed numerical study of the VSI for two classes of equilibrium disc structures with controversial results on the cooling criterion. 
The vertically isothermal models indeed demonstrated the requirement of very rapid thermal relaxation, $\Omega t \la 10^{-2}$, for the instability to develop. 
In contrast, the vertical polytropic models pushed the critical relaxation limit as far as to $\Omega t \approx 10$ for specific temperature and density profiles. 
In analytic investigation of the VSI including the radiative cooling, \citet{LIN15} confirmed a generic utility of the critical cooling timescale of the form 
\begin{equation}
    \Omega t_\mathrm{c} = \frac{h|q|}{\gamma - 1}
    \label{eq:tcrit}
,\end{equation}
with $h$ being the disc aspect ratio, $q$ the radial power-law temperature gradient, and $\gamma$ the adiabatic index. 
\citet{LIN15} showed that time \eqref{eq:tcrit} is a robust critical relaxation criterion for modes $5 \leq \Tilde k_r \leq 10$. 
For low-frequency modes $\Tilde k_r = 1$, the threshold is at slightly longer times. 
For high-frequency modes $\Tilde k_r \ga30$, the linear growth carries on for cooling times considerably longer than \eqref{eq:tcrit}, but the growth rate drops significantly \citep[see Fig.13 in][]{LIN15}.
In Fig.~\ref{fig:scan}, the zone of onset of VSI modes $\Tilde\lambda_r = 0{.}628$, predicted by $t < t_\mathrm{c} = 6{.}25\times10^{-2}\Omega^{-1}$, is hashed with grey lines. 
This occupies a sizeable fraction of the disc volume, and spans from $\approx15\mbox{ au}$ to $\approx180\mbox{ au}$ in the midplane. 
Above the midplane, the VSI can develop closer to the star. 

    \subsubsection{Convective overstability}\label{sec4:cov}
Jumps in opacity, surface density, ionisation, or stellar irradiation at close radii can introduce a negative radial entropy gradient, at least in some specific locations in a disc. 
This negative entropy gradient makes the protoplanetary discs baroclinic\footnote{Baroclinic means having misaligned isobars and isopycnals}, which provides another route to instability \citep{KLAHR01b,RAETTIG13}. 
Recently, \citep{KLAHR14} has shown that radially stratified baroclinic flows are linearly unstable to non-axisymmetric perturbations given a finite thermal relaxation time. 
During an oscillatory epicyclic motion, the parcel of gas heats up when displaced radially inwards, and cools down when displaced radially outwards. 
When crossing its starting position, the gas has a different temperature from its surroundings if the relaxation time is non-zero. 
The related density difference enhances the restoring buoyant acceleration (overstability), and the initial oscillation is amplified. 
This phenomenon obtained the name `convective overstability' \citep{KLAHR14}. 
The finite-amplitude noise resultant from the convective overstability (COV) can trigger a non-linear state of vortex amplification coined as subcritical baroclinic instability \citep[SBI,][]{KLAHR03a,PETERSEN07,LESUR10,LYRA11}. 
\citet{LYRA14} confirmed the phenomenon via compressible linear analysis and demonstrated in both 2D and 3D shearing box runs that the SBI is indeed the saturated state of the convective overstability.  
The lifetime of the resultant vortex is also a non-monotonic function of thermal relaxation rate.
The exact nonlinear outcome of the COV in discs with radial and vertical stratification is yet to be determined in numerical modelling \citep{LATTER16}. 
The SBI eventually yields growing vortices, which can induce Reynolds stresses rateable to $\alpha = 10^{-4} - 10^{-2}$ \citep{RAETTIG13,RAETTIG15,AIARA15}. 
These anticyclonic vortices can also be triggers for fast and efficient planetesimal formation \citep{KLAHR03b}. 

The growth rate of linear $r-z$ mode of the COV peaks at $\Omega t \gamma = 1$ \citep[where $\gamma$ is the gas adiabatic index,][]{KLAHR14}. 
Yet the growth rate of the non-linear regime, the self-similar amplification of vortices, peaks more likely at $\Omega t / \chi = 1$ ($\chi$ being the aspect ratio of the disc). 
Relaxation rates of $10^{-2} < \Omega t < 10^{2}$ are gained by $\Tilde\lambda_z = 2$ modes in the midplane from $\approx2\mbox{ au}$ to $\approx40\mbox{ au}$, and beyond $\approx130\mbox{ au}$. 
Away from the midplane, the suitable relaxation zones bend towards the central object (the light grey areas in Fig.~\ref{fig:scan}) forming two arc-shaped regions, where the COV can develop. 
The contour lines of $\Omega t = 1$ are approximately in the middle of each of those regions. 
For higher-frequency modes, $\Tilde\lambda < 2$, the inner arc region is even closer to the star: its inner edge is at $\approx0{.}6\mbox{ au}$ for $\Tilde\lambda=0{.}5$. 

    \subsubsection{Zombie vortices}\label{sec4:zvi}
\citet{MARCUS15,MARCUS16} showed that in 3D stably stratified baroclinic (Rayleigh-stable if unstratified) adiabatic shearing flows, an initial Kolmogorov turbulence undergoes a reverse energy cascade due to the background shear. 
In an environment with negative rotation shear (background Keplerian shear), anticyclonic regions merge to form a bigger anticyclonic vortex, while cyclonic regions get stretched into thin cyclonic vortex layers. 
At some point, wave solutions for vorticity become singular because of the shear, and a baroclinic critical layer is formed \citep{MASLOWE86}. 
This baroclinic critical layer then induces two vortex layers with the opposite sense of rotation. 
Anticyclonic layers are linearly unstable, roll-up into individual vortices, which then merge into one larger vortex \citep{MARCUS13}. 
The resultant vortex excites inertial gravity waves, which propagation is affected by the Coriolis force and the shear. 
The instability is the result of a resonant interaction between a Rossby wave and a gravity wave \citep{UMURHAN16}. 
Each generation of vortices excites new baroclinic critical layers, and the process repeats. 
The non-linear phase of self-reproduction of such structures obtained a name of zombie mode instability or zombie vortex instability (ZVI). 

The estimates of the angular momentum transport rate from the ZVI are not yet conclusive due to small local domains considered in numerical studies. 
The speculations are that the acoustic waves shed by zombie vortices can provide $\alpha\sim10^{-3}$ \citep{MARCUS15}.  
Also, the ZVI could contribute to turbulent mixing \citep[][]{TURNER07}. 

The inertial gravity waves, being driven by buoyancy, require long thermal relaxation times, $N^{2} t^{2}\gg1$ ($N$ is the Brunt-V\"ais\"al\"a frequency). 
\citet{LESUR16} found that the ZVI requires $\Omega t > 16$ and simultaneously Peclet numbers\footnote{a dimensionless number comparing nonlinear advection to thermal diffusion} of $\mbox{Pe} > 10^{4}$. 
This second condition constrains the volume, where the ZVI can operate, to the regions within $0{.}8\mbox{ au}$ in the radial direction and one local pressure scale height around the midplane (grey area in Fig.~\ref{fig:scan}). 
These zone is indeed MRI inactive according to \citet{DZYURKEVICH13}. 
Thus, the ZVI is a viable model for driving turbulence in inner regions of protoplanetary discs. 

\section{Discussion}\label{sec5:disc}
Hydrodynamic means of angular momentum transport in discs need to be studied because they can provide a turbulent stress in MRI-inactive regions and aid planetesimal formation. 
In numerical modelling for self-gravitating discs \citep{JOHNSON03,BOLEY06}, a usable estimate of the disc cooling results from vertically integrated azimuthally-averaged thermal energy balance \citep[see, however,][]{TAKAHASHI16}. 
In other identified hydrodynamic sources of disc turbulence, a resolution of the vertical direction and the high-frequency part of the perturbation spectrum is important \citep{NELSON13,KLAHR14}. 
For these, the cooling timescale of a disc annulus is not a viable estimate of the thermal relaxation \citep{NERO09,NERO10}. 
The relaxation timescale refers to a particular perturbation mode, and, in general, depends on its wavelength $\lambda$, location and orientation. 

In this work, we map radiative relaxation timescale of linear temperature perturbations of various wavelengths over disc interiors of given $\left( \rho,T \right)$ structures. 
We distinguish between two different regimes of relaxation: the optically thick and the optically thin (see Fig.~\ref{fig:ct}). 
Optically thick (in the Rosseland sense) perturbations relax on the radiative diffusion timescale, $t_\mathrm{thick}\propto\rho\tau_\mathrm{R}\lambda T^{-3}$. 
The optically thick relaxation is, in general, anisotropic. 
In the midplane, optically thick vertical modes $\lambda_z = 2H$ ($H$ is the pressure scale height) relax up to $60\%$ faster than radial modes $\lambda_r = 2H$. 
Above the midplane, the radial relaxation can be up to $100\%$ faster. 
Optically thin perturbations relax on the thermal emission timescale, $t_\mathrm{thin}\propto\kappa_\mathrm{P}^{-1}T^{-3}$, provided LTE. 
In our calculations, we account for the collisional decoupling from the emitters in the optically thin relaxation. Because of this, our relaxation timescales in low-density atmospheric layers and cold distant midplane regions are longer than those found by \citet{LIN15}. 
However, in the case of non-LTE, other radiative processes, not included in this study, should be taken into account, so the collisional timescale is, in general, not a reliable estimate of the relaxation timescale. 

The two relaxation regimes depend differently on relevant opacity: $t_\mathrm{thick}\propto\kappa_\mathrm{R}$, $t_\mathrm{thin}\propto\kappa_\mathrm{P}^{-1}$. 
The antipodal relation to opacity results in the opposite response to the reduction of dust amount \citep[][]{CAI06}.
We do not consider here any time evolution of the dust, which is required to accurately determine the relaxation timescales in the inner disc, as pointed out in \citet{NELSON00}. 
Our midplane temperatures, though, never exceed those of severe dust depletion, yet evaporation and condensation of the species is taken into account in the dust opacity tables \citep{SEMENOV03}.

We show that the diffusion timescales calculated in the flux-limited diffusion approximation \citep[as in][]{STOLL14} underestimate the relaxation timescale in the optically thin medium (see Fig.~\ref{fig:ct}). 
This is because the flux-limited diffusion formalism \eqref{eq:fld} does not account for the finite timescale of the intrinsic radiative processes but causal restriction on the energy density propagation (the free-streaming limit). 
A finite timescale of the collisional coupling \eqref{eq:tcoll} and finite ratio of the material's thermal capacity to its emissivity \eqref{eq:temit} both have to be taken into account. 
We report the bottom limit on the radiative relaxation of $\Omega t\approx10^{-4}$ (for the considered models) set by thermal emission timescale \eqref{eq:temit}. 

    \subsection{The method: thin and thick}\label{sec5:method}
In this section, we discuss the limits of the current method and highlight the measures for further improvement. 

In linear analysis, relaxation timescale \eqref{eq:trelax} is independent of the perturbation amplitude $\delta T_\mathrm{0}$ and is solely defined by the location within the disc. 
A value of the relaxation timescale can be assigned to each point $x$ of an optically thin Fourier mode $\delta T(\vec{x})$. 
The maximum of these values can be taken as the relaxation timescale of the mode.  
We make a simplification by taking the relaxation time at the point of extremal $\delta T = \delta T_\mathrm{min/max} = \delta T_\mathrm{m}$.  
This is a viable approximation given smooth Planck opacity within the perturbation wavelength. 

The optically thick relaxation timescale demands non-local calculations: the photons emitted at the point of $\delta T_\mathrm{0}$ subsequently diffuse through the medium with $0 < |\delta T / \delta T_\mathrm{m}| < 1$. 
Variations in the diffusion length scale at all those points alter the relaxation timescale. 
We use an effective diffusion coefficient harmonically averaged over the perturbation wavelength $\lambda$ (see Appendix~\ref{app:A2}). 
For an arbitrary optically thick perturbation $\left( k_z, k_r, m \right)$, the bulk of the excess thermal energy escapes through the direction of the least Rosseland depth \citep[cf. the flashlight effect, ][]{NAKANO89}. 
The optically thick relaxation is, therefore, sensitive to the disc density structure and especially its non-isotropy or time fluctuations. 
Numerical modelling with time evolution is required here. 

The radiative diffusion and thermal emission timescales both imply fast -- on timescales shorter than the relaxation timescale itself -- thermal coupling between the radiating (dust and gas) species and the thermal carriers (the gas). 
If thermal emission timescale \eqref{eq:temit} is shorter than the time to set LTE, we limit the relaxation by the appropriate collision timescale. 
Non-LTE cooling and non-equilibrium thermal feedback processes between the dust, the gas and the radiation field (external irradiation) are the necessary further refinements. 

The optical depths and emission rates are calculated using the joint dust and gas mean opacity tables obtained by direct summation of the corresponding mean values. 
At low temperatures ($\la1\,500\mbox{ K}$), the Rosseland opacity of the dust exceeds that of the solar-mixture gas by several orders of magnitude. 
Hence, adding the two means to each other does not introduce a considerable error far from the dust sublimation temperature: the opacity is either dominated by the dust or by the gas. 
Further, data from \citet{SEMENOV03} accounts for the freeze out of volatile organics, water and carbon monoxide, the main gas-phase absorbers at low temperatures. 
In the case of considerable (by a factor of $\approx10^{2}$ or more) dust depletion, this approach is invalid because the Rosseland averaging is not additive. 

Dust opacity depends not only on the dust chemistry but on the grain size distribution as well \citep{POLLACK94,HENNING96}. 
As pointed out in \citet{NELSON00}, the grain size evolution proceeds on short (relative to the local orbital period) timescales in the inner disc. 
Calculating the radiative relaxation there necessitates considering time-dependent size and composition of the grains.
In the atmospheric layers imposed to an external radiation field, the photochemistry breaks the assumptions of equilibrium chemistry and LTE, under which the opacities were calculated. 
The resolution of this issue requires running a frequency-dependent radiative transport on top of a non-equilibrium chemical reactions network and disc hydrodynamics. 
However, gas chemistry in dusty environments of discs is poorly constrained: many important reaction rates and transitions have not yet been measured in laboratories \citep{DUTREY14}. 
Our two constructed gas opacity models probe the parameter space. 
The analysis could benefit from the yet missing consistent dust and gas opacity calculations.  

\section{Summary}\label{sec6:summ} 
The radiative cooling timescale is important in the context of disc stability, associated transport, and disc observational appearance. 
A simple recipe to compute the radiative relaxation timescale in protoplanetary discs is introduced. 
The relaxation timescale as a function of perturbation wavelength, temperature and density is obtained by means of non-compressible linear analysis. 
There are two regimes of relaxation depending on how the thermal energy evolves: the optically thick and the optically thin.  
First is determined by radiative diffusion, second -- by thermal emission. 
In comparison to previous studies, collisional decoupling is taken into account in the optically thin relaxation, both the Planck and the Rosseland tabulated opacities of the dust and the gas are used, different dust depletion degrees are considered. 
The optically thick regime can be both dynamically fast and slow ($\Omega t \leogr 1$), the optically thin one is dynamically fast if not limited by collisional coupling. 
The fastest, LTE relaxation is $\Omega t \sim 10^{-4}$ in a typical T Tau disc. 

With this recipe, one can evaluate the relaxation timescale of a grid cell or an SPH particle based on its actual optical depth. 
This has applications for studying hydrodynamic and gravitational instabilities in the context of planet formation and transport in accretion discs, core migration, and planet-disc interaction. 
Calculated maps of the relaxation timescale constrain the origins of the identified hydrodynamic instabilities in discs (Fig.~\ref{fig:scan}). 
A large volume of a typical T Tau disc is unstable to the development of linear hydrodynamic instabilities: the convective overstability \citep[COV,][]{KLAHR14} and the vertical shear (a.k.a. Goldreich-Schubert-Fricke) instability \citep[VSI,][]{NELSON13}. 
The zone of operation of the VSI spans from $15\mbox{ au}$ to $180\mbox{ au}$ in the midplane. 
The COV can develop both closer in and farther out than that but there is a COV-dead layer from $40\mbox{ au}$ to $140\mbox{ au}$ in the midplane, where relaxation is too fast ($\Omega t < 10^{2}$) even for the low-frequency modes $\lambda = 2H$. 
A subcritical hydrodynamic instability, the zombie vortex replication \citep{MARCUS13}, can work inside $\approx0{.}8\mbox{ au}$ and within one pressure scale height, where thermal ionisation is yet inefficient to drive the MRI. 
Hydrodynamic modelling with time-evolution of the opacities is required for further study. 

\begin{acknowledgement}
    This research has been supported by the International Max-Planck Research School for Astronomy and Cosmic Physics at the University of
    Heidelberg (IMPRS-HD). M.M. thanks Hubert Klahr for valuable insights on hydrodynamic stability of rotating fluids, Kai-Martin Dittkrist for fruitful discussions and hints. 
    The authors thank the anonymous referee for useful comments on the scope of the presented approach and various suggestions, which significantly improved the appearance of the material. 
\end{acknowledgement}
\appendix

\section{Optically thick relaxation}\label{app:A}
    \subsection{Basic equations}\label{app:A1}
To obtain a characteristic damping time of optically thick thermal perturbations, we start with the thermal energy evolution equation considering only radiative processes \citep[see][]{KUIPER10a}
\begin{equation}
    \partial_t \left( E_\mathrm{int} + E_\mathrm{R} \right) = -\nabla \vec{F}_\mathrm{R} .
    \label{eq:enbalA}
\end{equation}
Here $E_\mathrm{int}$ is the internal energy density of the gas, $E_\mathrm{R}$ the radiation energy density, $\vec{F}_\mathrm{R}$ the flux of the radiation energy density. 
At high optical depths, the temperature of the gas and the ambient thermal radiation field are equal. 
The internal energy density of the gas with mass density $\rho$ and temperature $T$ is given by
\begin{equation}
    E_\mathrm{int} = \rho C_\mathrm{v} T
    \label{eq:eint}
,\end{equation}
with $C_\mathrm{v}$ being the specific thermal capacity at constant volume.
The radiation energy density is defined by
\begin{equation}
    E_\mathrm{R} = aT^{4},
    \label{eq:erad}
\end{equation}
where $a = 4\sigma/c$ is the radiation constant. 
Use of \eqref{eq:eint} and \eqref{eq:erad} in the left-hand side of \eqref{eq:enbalA} yields 
\begin{equation}
    \partial_t \left( E_\mathrm{int} + E_\mathrm{R} \right) 
    = 4 a T^{3} \left( 1 + \frac{\rho c_\mathrm{V}}{ 4 a T^{3}} \right) \partial_t T 
    = 4 a T^{3} \left( \frac{3}{4} + \frac{1}{4}\eta^{-1}  \right) \partial_t T
    \label{eq:lhs2t}
.\end{equation}
Here, $\eta = E_\mathrm{R} / (E_\mathrm{R} + E_\mathrm{int})$. 
The flux of the thermal radiation energy reads
\begin{equation}
    \vec{F}_\mathrm{R} = - D \nabla E_\mathrm{R}. 
    \label{eq:frad}
\end{equation} 
In the framework of the flux-limited diffusion, the diffusion coefficient $D$ is
\begin{equation}
    D = \frac{ \varkappa c }{ \kappa_\mathrm{R} \rho }
    \label{eq:fld}
,\end{equation} 
with $\kappa_\mathrm{R}$ being the Rosseland mean opacity, $c$ the speed of light, $\rho$ the mass density, and $\varkappa$ the flux limiter \citep{LEVERMORE81}. 
The gradient of the thermal radiation flux unveils
\begin{equation}
    -\nabla\vec{F}_\mathrm{R} = 
    4 a T^{3} D\times \left( \triangle T + \frac{3\left(\nabla T\right)^{2}}{T} + \vec{A} \cdot \nabla T \right)
    \label{eq:gradfrad}
,\end{equation} 
with 
\begin{equation}
    \vec{A} = \frac{\nabla \kappa_\mathrm{R}}{\kappa_\mathrm{R}} + \frac{\nabla\rho}{\rho} = \nabla \ln \kappa_\mathrm{R} + \nabla \ln \rho = \nabla \ln \rho \left[ 1 + \rho\partial_\rho\ln\kappa_\mathrm{R}\right]. 
    \label{eq:vecA}
\end{equation}

Equilibrium temperature $T_\mathrm{0}$ satisfies 
\begin{equation}
    4 a T_\mathrm{0}^{3} D\times \left( \triangle T_\mathrm{0} + \frac{3\left(\nabla T_\mathrm{0}\right)^{2}}{T_\mathrm{0}} + \vec{A} \cdot \nabla T_\mathrm{0} \right) = 0. 
    \label{eq:teq}
\end{equation}
Using \eqref{eq:lhs2t} and \eqref{eq:gradfrad} in \eqref{eq:enbalA}, one finds the temperature evolution equation
\begin{equation}
    4aT^{3} f^{-1} \partial_t T = 
    4 a T^{3} D\times \left( \triangle T + \frac{3\left(\nabla T\right)^{2}}{T} + \vec{A} \cdot \nabla T \right) ,
    \label{eq:Tevol}
\end{equation}
with 
\begin{equation}
    f = \partial_T E_\mathrm{R} / \partial_T \left( E_\mathrm{int} + E_\mathrm{R} \right) = 4\eta/\left(1+3\eta\right) 
    \label{eq:fratio}
,\end{equation}
following \citet{KUIPER10a}. 
 
    \subsection{Linear analysis}\label{app:A2}
Linearisation $T = T_\mathrm{0} + \delta T$ of~\eqref{eq:Tevol} with the use of \eqref{eq:teq} gives
\begin{equation}
    f^{-1} \delta \dot{T} = D\left( \triangle \delta T 
    - \frac{ 3\left( \nabla T_\mathrm{0} \right)^{2} }{T_\mathrm{0}} \frac{ \delta T }{ T_\mathrm{0} } 
    + \left( \frac{6 \nabla T_\mathrm{0}}{T_\mathrm{0}} + \vec{A} \right) \cdot \nabla \delta T \right). 
    \label{eq:Tperturb}
\end{equation}
Let
\begin{equation}
    k = \frac{2\pi}{\lambda}
    \label{eq:k}
,\end{equation}
be a wavenumber of the Fourier mode with spatial alignment along direction $x$ of the fastest radiative heat transport (not necessarily the vertical direction).
A dissipation time of this mode is then found by employing the Fourier ansatz $\delta T_{k} = \delta T_{k}^{0} \exp\left(-i\vec{k}\vec{x} + i\omega t\right) + c.c.$: 
\begin{equation}
    t_\mathrm{thick} = 
    \frac{\left( Df \right)^{-1}}{ k^{2} + 3\left(\nabla \ln T\right)^{2} - \vec{A}' \nabla T }.
    \label{eq:tthick1}
\end{equation}
Here $\vec{A}' = \partial_T \left[ \nabla\ln\rho \left( 1 + \rho\partial_\rho \ln\kappa_\mathrm{R}\right) \right]$ is a partial derivative of \eqref{eq:vecA} with respect to temperature, and
\begin{equation}
    \tilde{D} = Df 
    \label{eq:deff}
,\end{equation}
the effective diffusion coefficient. 
If the wavenumber is small (or the density fluctuations within it are significant: $\left| \nabla_z \rho \right| \rho^{-1} > k$), we perform averaging over $\lambda$ 
\begin{equation}
    t_\mathrm{thick} = \frac{\left< \left(\tilde{D} \right)^{-1} \right>} 
                            {  k^{2} + 3 \left< \left(\nabla \ln T \right)^{2}\right> 
                                        -\left< \vec{A}' \nabla T \right> }. 
    \label{eq:avtauT}
\end{equation}
Because the Rosseland mean for dust is a moderately varying function of density, $|\partial_\rho\kappa_\mathrm{R}| \ll \rho^{-1}\kappa_\mathrm{R}$, we approximate $\vec{A}' = \partial_T\nabla\ln\rho$. 
For an adiabatic equation of state, $\rho\propto T^{\gamma}$, $\vec{A}'\nabla T = -\gamma\left(\nabla\ln T \right)^{2} + \nabla\gamma\nabla\ln T$. 
Separating the terms proportional to first and second powers of $\nabla \ln T $ in the denominator, one finds 
\begin{equation}
    t_\mathrm{thick} = \frac{ \left( \tilde{D} \right)^{-1} }{ k^{2} + k^{2}_\mathrm{T} + k^{2}_\mathrm{A} }.
    \label{eq:tthick2}
\end{equation}
Here, $k_\mathrm{T} = \left(3+\gamma\right)^{1/2}\nabla \ln T$ is the equilibrium temperature wavenumber, and $k_\mathrm{A} = \left(\nabla\gamma\nabla\ln T\right)^{1/2}$ the adiabatic wavenumber.  
In a hydrostatic picture, we argue, the factor $\nabla\gamma$ makes $k_\mathrm{A}$ negligible in favour of the thermal wavenumber $k_\mathrm{T}$. 
The equation reduces to 
\begin{equation}
    t_\mathrm{thick} = \frac{\left( \tilde{D} \right)^{-1} }{ k^{2} + k^{2}_\mathrm{T} }.
    \label{eq:tthick3}
\end{equation}
The thermal wavenumber $k_\mathrm{T}$ is determined by the characteristic length scale of equilibrium temperature gradient. 
In a vertically isothermal radially stratified disc with $T\propto r^{-q}$ this term may be estimated as $\nabla T / T \sim -qh\mbox{H}^{-1}$, with $\mbox{H}$ being the local pressure scale height, and $h$ the local aspect ratio usually of the order of $0{.}05$. 
Thus, we set
\begin{equation}
    t_\mathrm{thick} = \frac{\left< \tilde{D} \right>^{-1}} { k^{2} }   = t_\mathrm{diff}. 
    \label{eq:tdiff}
\end{equation}
For high-frequency modes, $\lambda \ll \mbox{H}$, the equilibrium thermal wavenumber is negligible in favour of the wavenumber of the mode itself. 
For low-frequency modes, $\lambda \ga \mbox{H}$, \eqref{eq:tdiff} still holds true for geometrically thin discs (small aspect ratios). 
Hence, the optically thick relaxation time evaluates the diffusion time of the excess thermal photons over the mode wavelength. 
For the application cases of vertically isothermal disc, it is legitimate to permanently set
\begin{equation}
    \nabla T = 0.   
    \label{eq:zerogradt}
\end{equation}
In this case, the relaxation time of an optically thick mode depends on its wavenumber $k$, the radiative diffusion coefficient $D$, and the ratio $f$, which is smaller than unity in this regime (see Sect.~\ref{appA:EoT}). 

    \subsection{Energy or temperature evolution time?}\label{appA:EoT}
Dividing \eqref{eq:lhs2t} by the total thermal energy $E_\mathrm{tot} = E_\mathrm{R} + E_\mathrm{int}$ gives a relation between the energy and temperature evolution times
\begin{equation}
    t_\mathrm{E} = \left( 1 + 3\eta \right)^{-1} \times t_\mathrm{T}. 
    \label{eq:tauEtauT}
\end{equation}
If $\eta \ll 1$, the two times are equal up to $O(3\eta)$.
The ``critical'' density $\rho_\mathrm{cr} = aT^3/c_\mathrm{V}$ at which the radiation energy density equals the internal energy density is as low as
\begin{equation}
    \rho_\mathrm{cr} = 1{.}3\times10^{-16}\left(\frac{T}{100\mbox{ K}}\right)^{3} \mbox{ g cm}^{-3}. 
    \label{eq:rhocr}
\end{equation}
For instance, $\rho_\mathrm{cr} \approx 10^{-15}\mbox{ g cm}^{-3}$ at $300\mbox{ K}$, that is several orders of magnitude below the typical midplane values for a PPD in T Tau phase. 
We find $\eta$ between $10^{-5}$ and $10^{-3}$ within the optically thick regions in the fiducial disc model. 

\section{Optically thin relaxation}\label{app:B}
There are two important timescales in the optically thick relaxation: the timescale of the collisional coupling to the emitters, and the LTE emission timescale (see Sect.~\ref{sec2:tthin}).
    \subsection{Thermal emission time}\label{appB:themti}
Consider the following energy balance equation:
\begin{equation}
    C_\mathrm{v}\dot{T} = Q^{+} - Q^{-}.
    \label{eq:enbalB}
\end{equation}
Here, the terms $Q^{+/-}$ designate the specific (per unit mass) heating or cooling terms due to the perturbation on top of an equilibrium temperature field. 
The emission rate of matter at equilibrium temperature $T$ is proportional to the Planck mean opacity $\kappa_\mathrm{P}$ at that temperature (Kirchhoff's law). 
The perturbed internal energy then evolves as 
\begin{equation}
    C_\mathrm{v} \delta\dot{T} = - 2 \kappa_\mathrm{P} \sigma \left[ \left(T+\delta T\right)^{4} - \left( T - \delta T\right)^{4}\right], 
    \label{eq:de_evol}
\end{equation}
where $\delta T$ is the amplitude of the temperature perturbation. 
Substituting a Fourier mode $ (T - T_{0})_{k} = \delta T_{k} = \delta T_{k}^{0} \exp\left( -i\vec{k}\vec{x} -i\omega t \right)$, we linearise
\begin{equation}
    C_\mathrm{v}\delta\dot{T} = - 16 \kappa_\mathrm{P} \sigma T^{4} \frac{\delta T}{T}. 
    \label{eq:tthinlinear}
\end{equation}
This implies the evolutionary time
\begin{equation}
    t_\mathrm{emit} = \frac{C_\mathrm{v}}{16\kappa_\mathrm{P}\sigma T^{3}}
    \label{eq:temit0}
,\end{equation}
independent of the perturbation wavenumber. 

    \subsection{Collisional coupling}\label{appB:coll}
In PPDs, relative velocities in both dust-to-gas and gas-to-gas collisions are attributed to the thermal motions of the gas particles 
\begin{equation}
    v_\mathrm{coll} = v^\mathrm{g}_\mathrm{th} = \sqrt{\frac{3k_\mathrm{B}T}{\mu}},
    \label{eq:vcoll}
\end{equation}
where $\mu$ is the mean molecular weight, $k_\mathrm{B}$ the Boltzmann constant. 
The mean time between two subsequent collisions is 
\begin{equation}
    t_\mathrm{coll} = \frac{1}{ n\sigma_\mathrm{c} v_\mathrm{coll} },
    \label{eq:dgtcoll}
\end{equation}
where $\sigma_\mathrm{c}$ is the collisional cross section with the targets of number density $n$. 

        \subsubsection{Dust-to-gas collisions}\label{appB:dgcoll}
The underlying grain size distribution in \citet{SEMENOV03} is an MRN function \citep{MATHIS77} with a spectral slope of $3{.}5$ and the modified maximum grain size of $5\mbox{ }\mu\mbox{m}$ \citep{POLLACK85}:
\begin{equation}
    n\left( a \right) \propto a^{-3{.}5}, \mbox{ } a_\mathrm{-} = 0{.}005\mbox{ }\mu\mbox{m}, \mbox{ } a_\mathrm{+} = 5\mbox{ }\mu\mbox{m}
    \label{eq:mrn}
.\end{equation}
This results in the effective collisional cross section $\sigma\approx\pi a_\mathrm{+}a_\mathrm{-}\approx1{.}5\times10^{-9}\mbox{ cm}^{2}$. 
        \subsubsection{Gas-to-gas collisions at low temperatures}\label{appB:ggcoll}
The emissivity of the gas is supplied by different species at different temperatures. 
In LTE, the most efficiently radiating species can be found by comparing their relative contributions to the total Planck mean 
\begin{equation}
    w^i = \frac{\kappa^i_\mathrm{P}}{\kappa_\mathrm{P}}, \mbox{ } \sum_i w^i = 1.
    \label{eq:wi}
\end{equation} 
Here $i$ labels the species (e.g. CO, H$_2$O, TiO, NH$_3$, etc.), and 
\begin{equation}
    \kappa^i_\mathrm{P} = \frac{\int\kappa^i_\nu B_\nu\, d\nu}{\int B_\nu\, d\nu}
    \label{eq:kpi}
,\end{equation}
is a contribution to the Planck mean from specie $i$. 
The effective coupling time to radiative gas species can be estimated as
\begin{equation}
    t^\mathrm{g}_\mathrm{coll} = \frac{1}{ \sum_i w^i t^{-1}_i }.
    \label{eq:tgcoupl}
\end{equation}
At the temperatures of interest, the most abundant species, H$_2$ and He, are not efficient emitters in terms of \eqref{eq:kpi}, but they store the bulk of thermal energy. 
The rest of the matter can be modelled as a sort of efficient emitters satisfying 
\begin{equation}
    \kappa'_\mathrm{P} = \kappa_\mathrm{P}. 
    \label{eq:r1}
\end{equation}
Coupling timescale \eqref{eq:tgcoupl} is then approximated as collisional timescale
\begin{equation}
    t^\mathrm{g}_\mathrm{coll} = t'_\mathrm{coll} = \frac{1}{n'\sigma_\mathrm{nn} v_\mathrm{th}}
    \label{eq:r2}
,\end{equation}
with the effective neutral-neutral collisional cross section $\sigma_\mathrm{nn}$. 
We estimate 
\begin{equation}
    \begin{split}
        \Omega t^\mathrm{g}_\mathrm{coll} = 1{.}6\times10^{-6}%
        \left( \frac {f}                {10^{-2}}                       \right)^{-1}%
        \left( \frac {\sigma_\mathrm{nn}}       { 10^{-17}\mbox{ cm}^{2}}       \right)^{-1}\times \\
        \left( \frac {\rho}     {10^{-10}\mbox{ gcm}^{-3}}      \right)^{-1}%
        \left( \frac {T}                {100\mbox{ K}}  \right)^{-1/2}%
        \left( \frac {r}                {10\mbox{ au}}  \right)^{-3/2},
    \end{split}
    \label{eq:trcapp}
\end{equation}
where $f = n' / n_\mathrm{gas}$ is the number fraction of the radiating species. 
In our calculations, we put $f = 10^{-2}$ and $\sigma_\mathrm{nn} = 8{.}5\times10^{-17}\mbox{ cm}^{2}$.

\bibliographystyle{aa}
\bibliography{ThermalRelaxationTime}

\end{document}